\documentclass[seceq, preprint]{ptptex}

\usepackage{amsmath,amssymb}
\usepackage{bm}
\usepackage{graphicx}
\usepackage{subfigure}
\usepackage{verbatim}
\usepackage{wrapfig}
\usepackage{ascmac}
\usepackage{makeidx}
\usepackage[dvips]{color} 

\def\slashchar#1{\setbox0=\hbox{$#1$} % set a box for #1
\dimen0=\wd0 % and get its size
\setbox1=\hbox{/} \dimen1=\wd1 % get size of /
\ifdim\dimen0>\dimen1 % #1 is bigger
\rlap{\hbox to \dimen0{\hfil/\hfil}} % so center / in box
#1 % and print #1
\else % / is bigger
\rlap{\hbox to \dimen1{\hfil$#1$\hfil}} % so center #1
/ % and print /
\fi}

\def\psekibun{\int \frac{\mathrm{d}^3p}{(2\pi )^3}}
\def\nsum{\sum_n}
\def\ln {\mbox{ln}}
\def\tr {\mbox{tr}}
\def\exp {\mbox{exp}}

\def\bpi {\mbox{\boldmath{$\pi$}}}
\def\btau {\mbox{\boldmath{$\tau$}}}

\def\pq2 {((p+\frac{q}{2})^2-M_0^2)((p-\frac{q}{2})^2-M_0^2)}
\def\Ep {E_p}

\def\bq {\mbox{\boldmath{q}}}
\preprintnumber[4cm]{%<-- [..]: optional width of preprint # column.
UT-Komaba/12-14}%\\PTPTeX ver.0.8\\ August, 1997}
\title{%        %You can use \\ for explicit line-break.
Quark-Hadron Phase Transition\\  
in the PNJL model for interacting quarks
}

\author{%       %Use \scshape for the family name.
Kanako \textsc{Yamazaki}\footnote{GCOE Research Assistant at the Physics Department of the 
Graduate School of Science, University of Tokyo, Hongo.} and T. \textsc{Matsui} %\textsc{Familyname}%
}

\inst{%     %Affiliation, neglected when [addenda] or [errata].
Institute of Physics, University of Tokyo, Komaba
}

\abst{%This abstract is neglected when [addenda] or [errata].
We study the quark-hadron phase transition at finite temperature with zero 
net baryon density by the Nambu-Jona-Lasinio model for interacting quarks in uniform 
background temporal color gauge fields.  
At low temperatures, thermal quark excitations are suppressed by a small expectation value of the Polyakov loop which prohibits excitation modes besides color singlet states, on the other hand at high temperatures quarks and anti-quarks are allowed to be excited almost freely with the Polyakov loop close to unity.
Mesonic excitations are incorporated into the correlation energy as thermal fluctuations of auxiliary bosonic fields.
We show that at low temperatures these mesonic excitations dominate the pressure, while at high temperatures quarks and gluons instead dominate the pressure, mesonic excitations being absorbed in continuum of quark anti-quark excitations. We also show that there are two types of mesonic excitations, collective modes and non-collective modes, and only collective modes contribute to pressure at low temperatures.}

\begin{document}

 \maketitle

\section{Introduction}
Studying the phase diagram of quantum chromo-dynamics (QCD) \cite{Baym:1984,Pis:1984, Fukushima:2010bq} is one of the most fundamental problems in modern nuclear physics. 
One expects that  %QCD has a very rich phase structure.  
at low temperatures and low densities,  the chiral symmetry is broken spontaneously, 
and matter consists of interacting hadrons, where colored quarks and gluons are confined in individual hadron,  
while at sufficiently high temperatures and/or high baryon densities, the chiral symmetry is restored 
and matter is expected to turn into a plasma of deconfined quarks, antiquarks and gluons.
Such extreme states of matter, commonly called the quark-gluon plasma, have been
intensively studied experimentally by ultrarelativistic collisions of heavy nuclei with RHIC at Brookhaven and 
now with LHC at CERN\cite{QM:2011}. 

All QCD phase transitions are expected to occur in a non-perturbative regime with a strong QCD 
coupling.   
Although lattice QCD simulations by the Monte Carlo integration method have been shown very powerful in 
studying non-perturbative regime of QCD, it is plagued with the notorious sign problem in applying the same 
method to describe the region at finite baryon densities. 
For the exploration of the entire structure of the QCD phase diagram, we still need to resort to a more phenomenological approach adopting an effective theory which embodies some essential features of QCD.

The Nambu-Jona-Lasinio (NJL) model was originally designed to describe profound consequences of 
presumed underlying chiral symmetry in hadron dynamics in analogy to the BCS theory of superconductivity.\cite{NJL61}
It has later been transcribed to describe effective quark dynamics incorporating the symmetry 
%at the underlying quark level
\cite{HK94}.   
The model, however, has no mechanism of quark confinement:  
therefore, when applied in the mean field approximation to a finite temperature system,  it 
generates thermal excitations of quark quasiparticles even at low temperature, 
where quarks are to be confined in individual mesons and baryons.

To remedy this problem, a phenomenological model %which removes these unphysical modes of excitations 
was proposed by Fukushima\cite{Fuk04} by making use of the Polyakov loop 
%for a color charge nailed down at a fixed spatial point 
%, logarithm of whose expectation value 
%gives negative of free energy of single fundamental charge. \cite{Pol78,Sus79}
which is known to characterize confinement-deconfinement transition 
in strong coupling pure gauge theories.\cite{Pol78}
The Polyakov loop was inserted as a sort of fugacity factor in the quark distribution function 
and suppresses the quark single particle distribution by phase cancellations at low temperatures, 
leaving behind excitations of triad of quark quasiparticles in color singlet configuration. 
The quark-liberating deconfining transition was described as a change of the expectation value of 
the Polyakov loop treated as an order parameter with effective Landau-Ginzburg type free energy functional.   

Fukushima's original model has been reformulated as a mean field theory in uniform background temporal 
gauge field and has been studied extensively by others.\cite{RTW06, RRW07,RHRW08} 
This model, referred to as the PNJL model, allows to incorporate mesonic correlation energies beyond
mean field approximation.\cite{RHRW08,HABMNR07}
Such mesonic correlation energy had been studied previously in the original NJL model\cite{HKZ94,FB96}, 
in analogy to the classic example of the RPA correlation energy in degenerate electron plasma.\cite{GB57}
%Numerical computation with the model has been shown to reproduce the equation of 
%state calculated by the Monte Carlo simulation of the lattice gauge theory at small non-zero 
%chemical potential rather well. 
The model has been used to compute other properties of hadronic matter,
such as susceptibility at finite chemical potential\cite{SFR07}, 
thermodynamics in the regime of complex chemical potential\cite{SKKY08} 
and the effect of confinement in the study of QCD phase diagram\cite{PB12}.

%It has been shown that the model still allows, at low temperature, excitations of a triad of 
%quark quasi-particles which carry unit baryon number.   
%Mesonic excitations in such models are known to be contained in the correlation energy, as collective 
%excitations, similar to the classic example of plasma oscillation.\cite{GB57}

In this paper we reexamine the model in detail at zero chemical potential. 
We show that equation of state of a meson gas can be derived explicitly at low temperatures by the method 
of auxiliary fields which physically express effective meson fields build-up as a quark-antiquark 
bound states as in the original Nambu-Jona-Lasinio model.  
Similar results have been obtained earlier by the other groups by numerical computations of the 
mesonic correlation energies with slightly different models.\cite{Megias:2004hj,RRW07,RHRW08,Blaschke:2007np}
We will show explicitly that in the chiral limit the equation of state becomes that of a gas of mesons which 
are described as collective modes of excitations and evaluate the correction due to the composite nature 
of mesons as the individual excitations of the quark triad.  
It is found that the former dominates the equation of state at low temperature and the latter do not 
alter the free meson gas result significantly. 
As the temperature increases the mean field contribution from underlying quark quasiparticle 
excitations become important due to the non-vanishing Polyakov loop expectation value and will 
eventually dominate the free energy in deconfining phase.  
Collective mesonic excitations melt into the continuum of quark anti-quark excitations and the non-collective 
mesonic correlation remains negligible compared to the mean-field result. 
 
%{\bf Are these conclusions right? We need to check!}
The rest of this paper is organized as follows. 
In the next section, we present the NJL model  in uniform color gauge fields.  
We introduce bosonic fields as auxiliary fields in order to describe meson condensates and mesonic correlations 
and change four point NJL interactions to Yukawa type interactions.  
In section 3, the mean field approximation is reviewed. Under the mean field approximation, 
the fluctuation of meson fields are neglected. 
Therefore pressure is written in terms of constant auxiliary meson fields and thermalized quark quasiparticles. 
However these quark quasiparticles are not actually excited at low temperatures because the small expectation 
value of the Polyakov loop prohibits excitations of colored states. 
We discuss this mechanism in section 4 by focussing on the modification of the quark distribution function.  
In this section, we also note a change of the effective degrees of freedom between the low temperature phase 
and the high temperature phase. 
In section 5, we calculate the contribution of mesonic correlation to the equation of state and show that they are decomposed into collective meson excitations and non-collective individual excitations of quarks or quark-triads. 
Especially in the chiral limit, where the bare quark mass vanishes, the contribution of collective modes is separated from 
that of non-collective modes and gives the pressure of  gas of free pions and free sigma mesons. 
We present our numerical results in section 6. 
It is shown that the pressure of  free  pion gas is converted continuously to that of free quark-gluon gas as the temperature increases,
exhibiting a cross-over transition from hadron gas to a quark-gluon plasma. 
We also examine in detail the role of collective and non-collective degrees of freedom in low temperature confining phase and
how they change as temperature increases, both in the chiral limit with massless pions and in more realistic situation with 
non-zero pion mass, by studying the modification of the dispersion relations. 
It will be shown that the collective meson poles are absorbed into the continuum of quark-antiquark excitations at high 
temperatures and individual excitations do not contribute significantly to the pressure of the system at all temperatures.
We summarize this work in section 7. 
Some details of our computations are given in appendices.  
A condensed version of this work has been reported in \cite{YM2012}.

\section{NJL model in uniform color gauge fields}  
We begin with an Euclidean path-integral expression of the partition function for the NJL model 
of quarks in external temporal color gauge field:  
\begin{equation}
Z ( T, A_4 ) = \int [d q ][d \bar q ]  \exp{ \left[  \int_0^\beta d \tau  \int d^3 x 
{\cal L}_{\rm NJL} ( q, {\bar q}, A_4 ) \right] } 
\label{eq:NJL_PF}
\end{equation}
where the effective Lagrangian is given by %taken as
\begin{equation}
{\cal L}_{\rm NJL} ( q, {\bar q}, A_4 ) = {\bar q} ( i \gamma^\mu D_\mu  - m_0 ) q 
 + G \left[ ( {\bar q} q )^2 + (i{\bar q} \gamma_5 \tau q)^2 \right]  %- {\cal U} [ \Phi (A_4)] 
\end{equation}
for two flavor light quarks, $ {\bar q} = ( {\bar u}, {\bar d} ) $ in external (classical) temporal color gauge field, 
$D_\mu = \partial_\mu + g A_0  \delta_{\mu, 0} $ and $m_0$ is the bare quark mass which breaks the chiral 
symmetry explicitly.  Here we have used a standard Minkovsky metric notation with the real time %$t = x_0$
being replaced by the imaginary time $\tau = i x_0$ and the time component of the gauge field replaced by $A_4= i A_0$. 
\cite{KG06}
We note here that quark fields are the only dynamical variables which describes thermal excitations of the system, 
while the temporal component of SU(3) gauge fields $A_4 = \frac{1}{2} \lambda^a {\cal A}^a_4$,  
$\lambda^a$ being the 3 $\times$ 3 Gell-Mann matrices, only plays a side role 
imposing constraints on color configurations of thermal quark excitations.  
In the following calculation, we take Ansatz of diagonal representations for $A_4$ 
%= (e^{i \phi_1}, e^{i \phi_2}, e^{-i (\phi_1 + \phi_2)}) $ 
along with the earlier works.\cite{Fuk04,RTW06, RRW07,RHRW08}

This partition function, Eq.(\ref{eq:NJL_PF}), contains the forth power of quark fields 
so that  we cannot perform the integration over the fermion fields. For solving this problem, 
we introduce four auxiliary bosonic fields $\phi_i = (\sigma, \bpi) $ coupled 
to quark densities $({\bar q} q, i {\bar q} \gamma_5 \btau q)$ by multiplying $Z (T, A_4) $ by a constant Gaussian integral: 
\begin{equation}
 \int [d \phi ] \exp{ \left[ - \frac{1}{4G} \int_0^\beta d \tau   \int  d^3 x \left( (\sigma + 2G {\bar q} q -m_0)^2 +  (\bpi + 2G i {\bar q} \gamma_5 \btau q)^2 \right) \right] }
\end{equation}
where periodic boundary conditions are applied for each of the auxiliary fields in the imaginary temporal 
direction: $\phi_i (\tau) = \phi_i (0)$, and the integration measure is chosen so that the integral gives unity.
This procedure, known as the Hubbard-Stratonovich transformation\cite{St57, Hu59}, converts the four point NJL quark interaction to Yukawa coupling 
of the bosonic fields $\phi_i$ to corresponding four quark densities.  
\begin{equation}
Z ( T , A_4) = \int [d q ][d \bar q ] [d \phi] \exp{ \left[  \int_0^\beta d \tau  \int  d^3 x 
{\cal L}_{\rm eff} ( q, {\bar q}, \phi, A_4) \right] }
\end{equation}
where
\begin{equation}
{\cal L}_{\rm eff} (q, {\bar q}, \phi ,  A_4 ) = {\bar q} \left[ i \gamma^\mu D_\mu  + \sigma + i  \gamma_5 \btau \cdot \bpi \right] q 
- \frac{1}{4G} ( (\sigma - m_0)^2 +  \pi_i^2 ) . %- {\cal U} [ \Phi (A_4)] 
\end{equation}
%Here we have shifted the $\sigma$ field by the bare quark mass $\sigma = \sigma - m_0$ so that the explicit chiral symmetry breaking effect 
%appears in mesonic interactions.  
Performing the Grassmann integral over the quark Dirac fields, the partition function becomes
\begin{equation}
Z ( T, A_4 ) = \int [d \phi  ] e^{ - I ( \phi ,  A_4 ) }
\end{equation}
where the exponent is given by
\begin{equation}
I ( \phi ,  A_4 ) =   \frac{1}{4G} \int_0^\beta d \tau   \int d^3 x  \left( (\sigma - m_0)^2 + \bpi^2 \right)
 -  \rm{Tr}  \ln \left[ \beta \left( i \gamma^\mu D_\mu  + \sigma + i   \gamma_5 \btau \cdot \bpi \right)   \right]  
% + {\cal U} [ \Phi (A_4)] 
\label{Imeson}
\end{equation}
where the trace in the second term is taken over the arguments of quark fields, including the space-time coordinates with anti-periodic boundary condition 
in the imaginary time axis, Dirac gamma matrices and the isospin and color indices.  
For constant auxiliary meson fields ($( \sigma, \bpi) = $ constants),  the trace can be expressed as a sum over the fermionic (quark) Matsubara frequencies 
($\epsilon_n = (2n+1) \pi T$) and integrals over the spatial momenta: 
\begin{eqnarray}
I ( \phi  , A_4 ) & = & -  \frac{\beta V}{4G} \left( (\sigma - m_0)^2 + \bpi^2 \right)  \nonumber \\ 
& & \qquad +  2 V N_f \sum_n  \int \frac{d^3 p}{(2\pi)^3}  \tr_c 
\ln \left[ \beta^2 \left( ( \epsilon_n - g A_4)^2 + {\bf p}^2 +\sigma^2 + \bpi^2 \right)  \right]  
% \left[ i \gamma^\mu D_\mu  - \sigma + i   \gamma_5 \btau \cdot \bpi  \right]  
% + {\cal U} [ \Phi (A_4)] 
%\nonumber \\
%& & \qquad \qquad {\rm for}  \qquad \phi  = (\sigma, \bpi ) = {\rm const. }
\label{Lmeson}
\end{eqnarray}
leaving behind only the trace over the color indecies.  Here we have taken flavor sum with $N_f =2$ assuming degeneracy of up and down quarks 
($m_u = m_d=m_0$). 

We evaluate the functional integral over the mesonic auxiliary fields by the method of steepest descent.   
Let $\phi _0 = (\sigma_0, \bpi_0) $ give local minimum value of the integrand so that
\begin{equation}
 \left. \frac{\delta I }{\delta \phi _i}  \right|_{\phi  =  \phi _0} = 0  .
 \label{stationary}
\end{equation}
Hereafter we choose $\bpi_0 = 0$, along with $\sigma_0 = - M_0$, which evidently satisfy stationary conditions for effective pion fields.
Shifting the integration variables by, $ \varphi_i = \phi _i  - \phi _0$, we expand the integral in a power series of $\varphi $: 
\begin{equation}
I ( \phi ,  A_4 ) =  I_0 % + \left. \frac{\delta I}{\delta \phi _i} \right|_{\phi  = \phi _0} \delta \phi _1 
+ \frac{1}{2} \left. \frac{\delta^2 I}{\delta \phi _i \delta \phi _j } \right|_{\phi = \phi _0} \varphi _i \varphi _j  
+ \cdots 
\label{eq:expansion}
\end{equation}
The first order term of  $\varphi $ becomes zero because of equation deciding local minimum point.
Keeping only up to quadratic term in the expansion, 
\begin{equation}
Z (T, A_4) \simeq e^{- I_0} \int [ d \varphi ] 
\exp{ \left[ - \frac{1}{2} \left. \frac{\delta^2 I}{\delta \phi _i \delta \phi _j } \right|_{\phi = \phi _0} 
\varphi_i \varphi_j  \right] }
\end{equation}
The gaussian integrals can be easily performed, yielding 
\begin{equation}
Z(T, A_4) = e^{-I_0} \bigl[ \mbox{Det}\ \frac{\delta I}{\delta \phi _i\delta \phi _j}\bigr]  ^{-\frac{1}{2}}
= e^{-I_0 - \frac{1}{2} {\rm Tr_M} \ln \frac{\delta I}{\delta \phi _i\delta \phi _j}  }
\end{equation}
which gives the thermodynamic potential:
\begin{equation}
\Omega (T,A_4) = T \left( I_0 +\frac{1}{2} {\rm Tr_M} \ln \frac{\delta I}{\delta \phi _i\delta \phi _j} \right) 
\label{eq:omega}
\end{equation}
up to one-loop fluctuations of the collective meson fields, where the trace is to be performed over the space-time coordinates of the auxiliary meson fields which obey a periodic boundary condition in the imaginary time 
direction. 

The first term of Eq.(\ref{eq:omega}) represents the thermodynamic potential under the mean field approximation and the second term represents the contribution of mesonic correlation to the thermodynamic potential. These mesons are free because we  kept the expansion of 
$I(\phi , A_4)$ up to quadratic term in Eq.(\ref{eq:expansion}). 
If we wish to take into account mesonic interactions, 
we need to keep higher order terms of expansion of $I(\phi , A_4)$. In this case we cannot integrate the partition function because of terms more than third power of $\varphi $ so that we have to calculate the partition function perturbatively. In this paper, we ignore interactions between mesons.

\section{Mean-field approximation}
The thermodynamic potential in the mean field approximation, $\Omega_{\rm MF} (T,A_4)$, or the corresponding 
pressure $p_{\rm MF}  (T,A_4)$, may be retrieved from the leading term $I_0$ by the relation: 
\begin{equation*}
\Omega_{\rm MF} (T,A_4) = T I_0  = - p_{MF} V .
\end{equation*}
The explicit form of the leading term $I_0$ is given by 
%\begin{equation}
%I_0 
%= \beta V \left[ \frac{1}{G} \left( (\sigma_0 - m_0)^2 + \pi_i^2 \right) %+ {\cal U} [ \Phi (A_4)]  
% \right] 
%+ V \sum_n \int \frac{d^3 p}{(2\pi)^3} {\rm tr} \ln \left[ \gamma_0 (\omega_n - g  A_4) + \gamma_i p_i - \sigma_0 \right]  
%\end{equation}
\begin{equation}
I_0 = 
   \beta V  \frac{1}{4G} (M_0 - m_0)^2 %+ {\cal U} [ \Phi (A_4)]   \right] 
-  2 V N_f \sum_n \int \frac{d^3 p}{(2\pi)^3}  \tr_c \ln \left[ \beta^2 \left( ( \epsilon_n - g A_4)^2 + {\bf p}^2 +M_0^2 \right) \right]  .
\end{equation}
%where we have taken the trace over the Dirac gamma matrices and the sum is over the Matsubara 
%frequencies $\epsilon_n = (2n+1) \pi T$ for quark fields which satisfy anti-periodic boundary conditions 
%in $tau$ and $\tr_c$ is the trace over the color indices.
This leads to
%where $\langle \cdots \rangle$ implies a statistical average over the external gauge field.
%Hence,
\begin{equation}
p_{\rm MF} (T, A_4) =   - \frac{1}{4G}  (M_0 - m_0)^2 + 
2 T N_f \sum_n  \int \frac{d^3 p}{(2\pi)^3}  \tr_c \ln \left[ \beta^2 \left(( \epsilon_n -  g A_4)^2 + {\bf p}^2 +{M_0}^2 \right) \right]  
\end{equation}
%where the trace is to be performed over the 3$\times$3 color matrix $A_4$. 
where $\epsilon_n$ is fermionic Matsuibara frequencies, $\epsilon_n=(2n+1)\pi T$,
the trace is to be performed over the 3$\times$3 color matrix $A_4$. The sum is taken over the Matsubara frequencies.
Evaluating the discrete sum over the Matsubara frequencies by the standard method of contour 
integration\cite{FW71}, we find
%The sum of the quark Matsubara frequency $\omega_n$ in the second term may be performed by the standard 
%method of contour integration in the complex $z$-plane\cite{FW71}: 
%\begin{equation}
%T \sum_n  \ln \left[ (\omega_n -  i g A_4)^2 + p_i^2 +{\sigma_0}^2 \right]  =
%\frac{1}{2 \pi i} \int_C d z \ln \left[ (z -  i g A_4)^2 + p_i^2 +{\sigma_0}^2 \right] f(z)
%\end{equation}
%with the multiplication factor $f (z) = 1/ (e^{\beta z} +1)$ which has a pole 
%at each Matsubara frequency $z = i \omega_n$ on the imaginary $z$-axis. 
%Changing the integration contour to the paths along the real $z$ axis, the sum of the infinite number of poles 
%on the imaginary $z$-axix may be converted to the sum of the residues of the poles on the real $z$-axis. 
%This gives,
\begin{equation}
p_{\rm MF} (T, A_4) =  p_{\rm MF}^0 (M_0) +  
2 N_f T  \int \frac{d^3 p}{(2\pi)^3} \tr_c \left[  \ln \left( 1 + e^{- \beta ( E_p + i g A_4) }  \right) 
+ \ln \left( 1 + e^{- \beta ( E_p - i g A_4) }  \right) \right]
\label{pmf}
\end{equation}
where
\begin{eqnarray}
p_{\rm MF}^0 (M_0) %& = &   \frac{1}{2G} (M_0 - m_0)^2 -  \int^{\Lambda}  \frac{d^3 p}{(2\pi)^3} \tr_c (E_p + igA_4) \nonumber\\
& =  &   3 \times 2 N_f  \int^{\Lambda}  \frac{d^3 p}{(2\pi)^3} E_p - \frac{1}{4G} (M_0 - m_0)^2 
\label{condpress}
\end{eqnarray}
is the pressure exerted by the zero point motion of the quark quasiparticles with energy 
$E_p = \sqrt{p^2 + M_0^2}$ in the "Dirac sea". 
The factor 3 in the first term of  (\ref{condpress}) accounts for color and the factor 2 for spin degeneracies. 
The second term in (\ref{condpress}) is the pressure due to the chiral condensate which shifts the quark 
mass from $m_0$ to $M_0$ and hence reduces the Dirac sea pressure. 
%The positive contribution of the chiral condensate pressure reduces the effect of the quark Dirac sea by the mass shift.  
This term may be understood as the subtraction of the double counting of the mean-field effect as in the Hartree approximation. 
%The contribution of the gauge field cancels out by the traceless condition $\tr_c A_4 = 0$. 
Note that this "vacuum pressure" does not depend on the external background gauge potential due to the cancellation of the effects of the potential on particle and antiparticle. 
It does not depend on temperature directly, but indirectly through the temperature dependence of the effective quark mass $M_0$.  
The last term in (\ref{pmf}) is equal to the entropy of the system carried by the thermally excitations of quasiparticles multiplied 
by the temperature.
It may be further reduced to more familiar form by partial integration, 
 \begin{equation}
p_{\rm MF} (T, A_4) = p_{\rm MF}^0  (M_0) + 
%4  \int \frac{d^3 p}{(2\pi)^3} \frac{p^2}{3 E_p}  \tr_c \left[ \frac{1}{e^{\beta ( E_p + i g A_4) }+1 } \right] 
2 N_f  \int \frac{d^3 p}{(2\pi)^3} \frac{p^2}{3 E_p}  \tr_c \left[ f ( E_p + i g A_4 ) +f ( E_p - i g A_4 ) \right]
\label{mf_pressure}
\end{equation}
where 
\begin{equation}
f ( E_p \pm i g A_4 ) = \frac{1}{e^{\beta ( E_p \pm i g A_4) }+1 } 
\label{qspdf}
\end{equation}
is  the quark (anti-quark) quasiparticle distribution function in the external gauge field potential. 
Note the change of sign in the gauge potential for quark ($igA_4$) and for anti-quark ($-igA_4$).  
The second term in (\ref{mf_pressure}) corresponds to the pressure exerted by quark and anti-quark quasiparticles
 in the external gauge potential; the factor 2 accounts for spin degrees of freedom. 
%As we expected, $A_4$ dependence appears only through the "Polyakov loop" phase factor, $L= e^{i g \beta A_4}$.  
The first term in  (\ref{pmf}) exists even in vacuum at zero temperature so that it will be removed when we readjust the zero of
the pressure.
The second term present the pressure exerted by thermally excited quark quasi-particles which carries entropy.
This term should be also removed in the low temperature phase where quarks are confined in hadrons.
This can be done by choosing a specific form of $A_4$ as will be described in the next section.   
%mimicking statistical average over the external color gauge field $A_4$.

In the mean field approximation, the value of the quark quasiparticle mass $M_0$ is determined by 
the condition that the term linear in the fluctuation $\delta \phi $ vanishes owing to the stationary condition (\ref{stationary})
which may be written explicitly, for $\sigma_0 = M_0$, 
\begin{equation}
M_0 - m_0 %= 2G N_f {\rm tr} \nsum \psekibun \frac{-1}{\slashchar{p}-M_0-i\gamma ^0A_4} 
= 8 G N_f  \nsum \psekibun \frac{M_0}{(\epsilon_n - g  A_4)^2+ p^2 + M_0^2} 
\label{gap}
\end{equation}
which is reduced to
\begin{equation}
M_0 - m_0 %= 2G N_f {\rm tr} \nsum \psekibun \frac{-1}{\slashchar{p}-M_0-i\gamma ^0A_4} 
= 8 G N_f  \tr_c \psekibun \frac{M_0}{E_p} \left( 1 - f ( E_p + i g A_4 ) - f ( E_p - i g A_4 ) \right) . 
\label{GE}
\end{equation}
%In the chiral limit $m_0 = 0$, this equation possesses a non trivial solution $M_0 \neq 0$ at low temperatures.  

\section{%Polyakov loop and 
Quenched quark distributions} 
In the above calculations, the constant temporal gauge field $A_4$ appears as a phase factor together with the quark quasiparticle energy in the single particle distribution function (\ref{qspdf}).  
It looks very similar to the gauge invariant Polyakov loop phase integral, 
\begin{equation}
L ({\bf r} ) = P \exp \left[   i g \int_0^\beta d \tau A_4 ({\bf r}, \tau )  \right] %=  e^{ i g \beta  A_4} 
\end{equation}
%In the PNJL model, $L$ 
whose thermal expectation value measures an extra free energy associated with the color charge in fundamental representation fixed at a spatial point $\bf r$.\cite{Pol78, McLerran:1981pb} 
Although this connection was first emphasized in the construction of the model by Fukushima, it looses its 
strict meaning in the present method, which follows the work of TUM group, since we assume that quarks are moving in a uniform background gauge field not fluctuating either in space or in imaginary time. 
Nevertheless we adopt the procedure of Fukushima, along with the TUM group, to replace the phase factor 
in the quark quasiparticle distribution function by the thermal average of the Polyakov loop.%: for the quark distribution function,  
%\begin{equation}
%\tr_c f ( E_p + i g A_4 ) =  \tr_c \frac{1}{  e^{\beta ( E_p + i g A_4) } +1 }   \to  \langle \tr_c \frac{1}{  L e^{\beta E_p  } +1 }  \rangle
%\end{equation}
Taking diagonal representation $L = (e^{\phi_1}, e^{i \phi_2}, e^{-i (\phi_1 + \phi_2)} )$,  one finds\cite{RHRW08}
\begin{equation}
\langle \frac{1}{3} \tr_c  f ( E_p + i g A_4 )  \rangle \to
% = \frac{(\bar{\Phi }+2\Phi e^{-\beta \Ep })e^{-\beta \Ep }+e^{-3\beta \Ep }}{1+3(\bar{\Phi }+\Phi e^{-\beta \Ep })e^{-\beta \Ep }+e^{-3\beta \Ep }}
\frac{ \bar{\Phi } e^{2\beta \Ep } + 2\Phi e^{\beta \Ep }+ 1}
{e^{3\beta \Ep } +3\bar{\Phi } e^{2 \beta \Ep } + 3 \Phi e^{\beta \Ep }+ 1}
 \equiv  f_\Phi ( \Ep)  
\label{averagedis}
\end{equation}
where 
\begin{equation}
\Phi = \frac{1}{3}  \langle \tr_c L  \rangle 
= \frac{1}{3}  \langle \left( e^{i\phi_1} + e^{i\phi_2} + e^{-i(\phi_1 + \phi_2)} \right)  \rangle ,  
\qquad \bar{\Phi} = \frac{1}{3}  \langle \tr_c L^\dagger  \rangle ~.
\end{equation}  
The anti-quark distribution, $\bar{f}_\Phi ( \Ep ) \equiv  \langle \frac{1}{3}\tr_c f ( E_p - i g A_4 )  \rangle $, 
is obtained from the quark distribution by the replacement of $\Phi$ by $\bar{\Phi}$ and vice versa. 
At zero chemical potential, two distributions are identical and $\Phi$ is real so that $\Phi = \bar{\Phi}$.
%is the trace of the expectation value of the Polyakov loop. 
We note this replacement corresponds to a Gaussian approximation in the statistical average, ignoring statistical correlations between the Poliakov loops, in the same spirit in the mean-field approximation.
%\footnote{We thank Gordon Baym and Tetsuo Hatsuda for discussion on this point.}

Very interesting observation here is that although in the deconfining phase where $\Phi = \bar{\Phi} =1$, one would simply have ordinary quark distribution, 
\begin{equation}
\left.  f _\Phi ( \Ep) \right|_{\Phi =1}  = \frac{1}{ e^{\beta E_p}+1 } ,
\end{equation}
%implying quark distribution for each of 3 colored state, 
in the confining phase, where $\Phi = \bar{\Phi} = 0$, we have instead, %this statistical average over the color gauge field yields,
\begin{equation}
\left.  f _\Phi ( \Ep) \right|_{\Phi =0} % =  f ( \Ep; 0)  
= \frac{1}{ e^{3\beta E_p}+1 } 
\label{triad}
\end{equation}
which may be interpreted as triad of three quark quasiparticles exciting together in color singlet configuration. 
It looks something like a baryon although no effect of interaction is taken into account between three quarks to
form a baryon, as reflected by the same value of momenta for all three quarks. 
We note that the degrees of freedom of excitations is not reduced to 1/3 of color triplet free quark excitations. However, there still exist reduction of effective degrees of freedom due to the change of the phase space of quark momentum $p$ and that of quark triad momentum $3p$:
The "triad" carries the energy $E_{\rm tri} = 3E_p$, the momentum $p_{\rm  tri} = 3p$ and the mass 
$M_{\rm  tri} = 3M_0$.
This implies that the Lorentz invariant phase space integral over the quark momentum is reduced when written in term of the triad momentum, e. g. 
%We find that the degree of freedom of "triad" becomes 1/9 of the degree of freedom of
%quark by changing the variable of integration from momentum of quark to momentum of "triad", e.g.
\begin{eqnarray}
\int \frac{d^3p}{E_p}\rightarrow \frac{1}{9}\int \frac{d^3p_{\rm tri}}{E_{\rm tri}}
\end{eqnarray}
%where $p'=3p$, $(p, E_p)\rightarrow (3p, 3E_p)$.

Here we determine $\Phi$ phenomenologically by adding an effective potential of $\Phi$:
\begin{equation}
\Omega (T, \Phi ) = \langle \Omega (T, A_4) \rangle + {\cal U} ( T,  \Phi ) 
\end{equation}
where we define, following\cite{RHRW08},
%the following effective potential is introduced % ${\cal U} [ \Phi] $ :
\begin{equation}
{\cal U} (T, \Phi ) / T^4 =  - \frac{1}{2} b_2 (T) \bar{ \Phi }\Phi - \frac{1}{6} b_3 ( \Phi^3 + \bar{\Phi}^3 ) + \frac{1}{4} b_4  (\bar{\Phi } \Phi )^2   
\label{LG-potential}
\end{equation}
with 
\begin{equation}
b_2 (T) = a_0+a_1\Big( \frac{T_0}{T}\Big) +a_2\Big( \frac{T_0}{T}\Big) ^2+a_3\Big( \frac{T_0}{T}\Big) ^3
\end{equation}
The parameters of the effective potential are determined so that it gives prescribed values for $\Phi$ at given temperature 
in analogy to the free energy of the order parameter in the Ginzburg-Landau theory of superconductivity.  
The temperature dependent parameters in the above expression of ${\cal U} [ \Phi] $ are chosen so that $\Phi =0$ for the low 
temperature confining regime where thermal quark quasi-particle excitations are forbidden,  while $\Phi \simeq 1 $ above 
deconfining temperature, quite opposite to the normal behavior of the order parameter: therefore 
the physical meaning of $\Phi$ becomes clearer if it is called a "disorder parameter".

We note here that, although gluon fields are not treated as dynamical variables, gluon excitations may be included
in a phenomenological fashion through the effective potential by setting,
\begin{equation}
{\cal U} (T, \Phi=1) =  \left( - a_0 - \frac{1}{3}b_3 + \frac{1}{4} b_4 \right) T^4 = - \frac{16 \pi^2}{90} T^4
\end{equation}
so that the equation of state matches to that of a free quark-gluon plasma at asymptotic high temperatures.

With all these prescriptions, the equation of state in the mean-field approximation of the PNJL model is 
given as a maximum of 
\begin{equation}
p_{\rm MF} (T, \Phi, M_0) =  p_{\rm MF}^0 (M_0) - \Delta p_{\rm vac}
+ 4  \times 3 \times \int \frac{d^3 p}{(2\pi)^3} \frac{p^2}{3 E_p}  f_\Phi ( E_p )  -  {\cal U} (T, \Phi) 
\label{mfpressure}
\end{equation}
with respect to variation of $\Phi$ and the quark quasiparticle mass $M_0$, which measures the magnitude 
of chiral condensate, the latter procedure gives the gap equation
\begin{equation}
M_0 - m_0 %= 2G N_f {\rm tr} \nsum \psekibun \frac{-1}{\slashchar{p}-M_0-i\gamma ^0A_4} 
= 12 G N_f \Bigl[ \int^{\Lambda}  \frac{d^3 p}{(2\pi)^3} \frac{M_0}{E_p}- \int  \frac{d^3 p}{(2\pi)^3} \frac{M_0}{E_p}\left(  f_\Phi ( E_p ) + \bar{f}_\Phi ( E_p )  \right) \Bigr] .
\label{gap0}
\end{equation}
which determines the optimal value of $M_0$.
The constant $ \Delta p_{\rm vac}$ is chosen so that the pressure becomes zero at zero temperature.  
In the confining phase, this equation could be interpreted as a gap equation for "baryon" (quark triad) mass,
\begin{eqnarray}
M_{tri} - 3m_0 = 3\times 12GN_f\times \frac{1}{27} \int \frac{d^3p_{tri}}{(2\pi )^3}\frac{M_{tri}}{E_{tri}}(1-f (E_{tri})-\bar{f} (E_{tri})).
\end{eqnarray}
If we use a rescaled coupling $G' = G/3$, this equation indeed looks more like original Nambu-Jona-Lasinio gap equation for baryons. 
This interpretation however meets difficulty when we try to interpret the quark quasiparticle pressure (\ref{mfpressure})
as a pressure from thermal excitations of baryon and anti-baryons.  
By rescaling the momentum of quarks, the second term of (\ref{mfpressure}) may be rewritten in the limit $\Phi = 0$ as 
\begin{equation}
4  \times 3 \times \frac{1}{3^4} \int \frac{d^3 p_{tri}}{(2\pi)^3} \frac{p_{tri}^2}{3 E_{tri}}  f ( E_{tri} ) 
\end{equation}
The reduction of the degrees of freedom is by the factor 1/27, more than what we need for the compensation of the remaining color factor 3. 
%\footnote{We thank Gordon Baym and Tetsuo Hatsuda for illuminating discussion on this subtle point.}

We show the temperature dependence of $\Phi$ and $M_0$ obtained from these mean-field equations in Fig. 1.  
In the chiral limit $m_0 = 0$, the "gap equation" (\ref{gap0}) possesses a non-trivial solution $M_0 \neq 0$ 
only at temperatures below a critical temperature $T_c$. 
In the ordinary NJL model, with no "Polyakov loop prescription", the chiral transition takes place at relatively low temperature,
while the confinement-deconfinement transition takes place at higher temperature as a first order transition with the discontinuous 
change in the value of $\Phi$.  
In the PNJL model, these two distinct transitions interfere each other through the quark quasiparticle loops.  
As a result, the two transitions takes place at a similar temperature:  the transition temperature for deconfinement becomes lower and becomes smooth cross-over transition, while the chiral transition takes place at higher temperature, since the quark excitations are suppressed at low temperatures by small expectation value of the Polyakov loop.    
This important observation was first made by Fukushima.\cite{Fuk04}
With a finite bare quark mass $m_0$,  the chiral symmetry is broken explicitly and the chiral transition also become a smooth cross-over transition. 

\begin{figure}[htbp]
\begin{center}
%\resizebox*{!}{4.7cm}{
 \includegraphics[clip,width=55mm, angle=270]{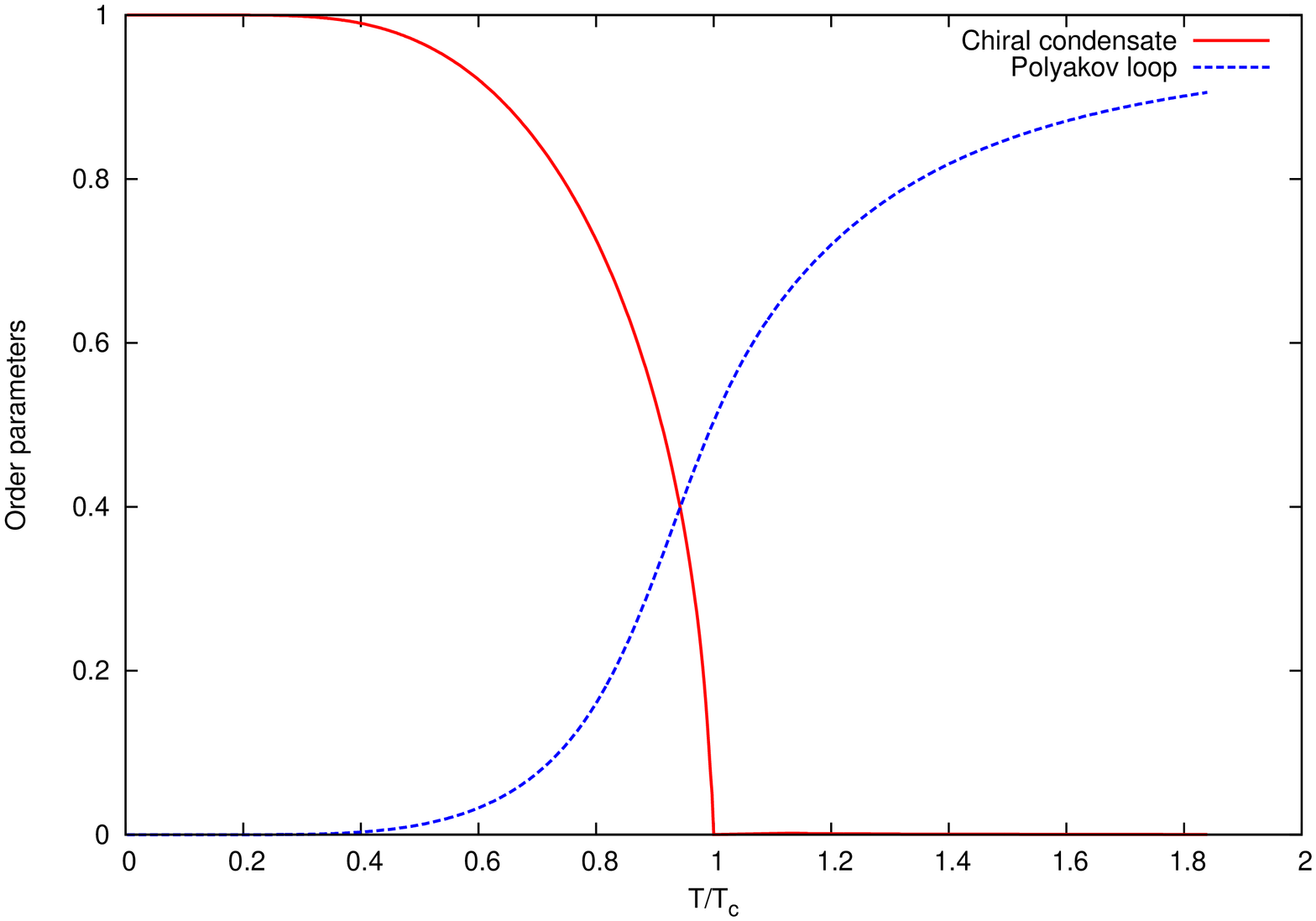}
 \includegraphics[clip,width=55mm, angle=270]{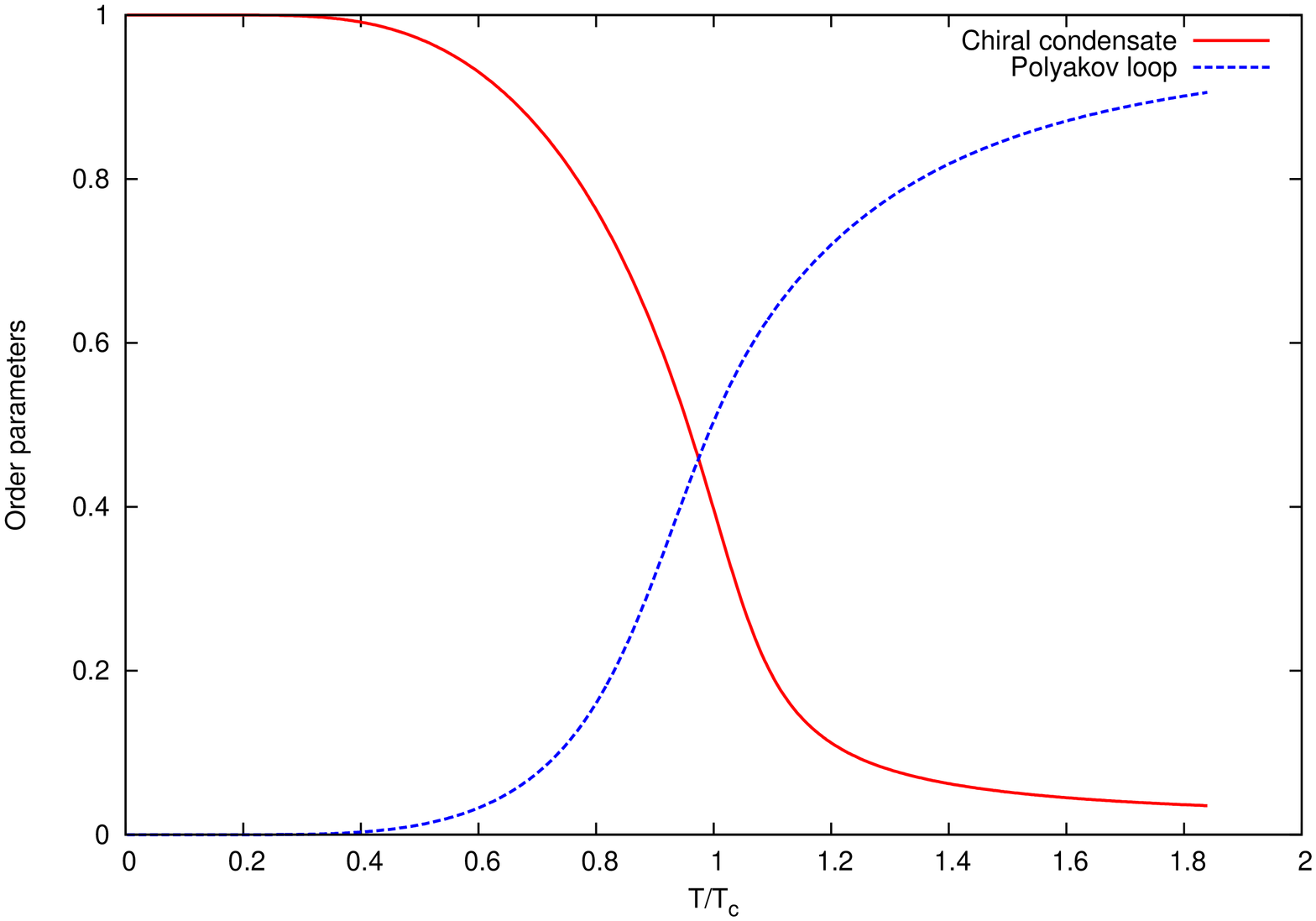}
\end{center}
\caption{
Temperature dependence of effective quark mass $M_0$ and expectation value of the Polyakov loop:
the left panel is for vanishing bare quark mass (the chiral limit) and the right panel is with finite bare quark mass $m_0$ which is chosen to reproduce the pion mass $m_\pi = 140$MeV.  
}
\end{figure}

\section{Mesonic correlations: collective v.s. non-collective}

Now we come to our main task to evaluate mesonic correlation energy.    
%This is done by performing the functional integral over the fluctuations in the auxiliary fields, $\varphi_i$, from their uniform mean values.   
%Retaining the quadratic term in the expansion of $I$, the integral becomes ordinary Gaussian integrals which 
%can be easily performed, yielding 
%\begin{equation}
%Z(T, A_4) = e^{-I_0} \bigl[ \mbox{Det}\ \frac{\delta I}{\delta \phi _i\delta \phi _j}\bigr]  ^{-\frac{1}{2}}
%= e^{-I_0 - \frac{1}{2} {\rm Tr} \ln \frac{\delta I}{\delta \phi _i\delta \phi _j}  }
%\end{equation}
The contribution of the mesonic correlation energy to free energy, $\Omega_M$, 
or the corresponding pressure $p_M = - \Omega_M/V$,  is given by
\begin{equation}
\Omega_M (T, A_4) %=   T  \ln Z(T, A_4) -I_0\rangle 
=  \frac{T}{2}  {\rm Tr}_M \ln  
\left. \frac{\delta^2 I}{\delta \phi _i \delta \phi _j } \right|_{\phi = \phi _0} 
\label{Momega}
\end{equation}
and
\begin{eqnarray}
p_M (T, A_4) %& = & - \Omega_M/V \nonumber \\
%& = & T \frac{1}{V} \langle  \ln Z(T, A_4) - I_0  \rangle \nonumber \\
& = & 
- \frac{T}{2V} {\rm Tr}_M \ln  
\left. \frac{\delta^2 I}{\delta \phi _i \delta \phi _j } \right|_{\phi = \phi _0} 
%\frac{1}{2} T \sum_n \int \frac{d^3 q}{(2\pi)^3} \ln \left[
% 2T\nsum \psekibun \frac{q^2}{\left[(\omega_n-iA_4)^2-E_p^2\right] \cdot \left[(\omega_n-iA_4+q_0)^2-E_{p+q}^2\right]} \right] \nonumber \\
\label{Mpressure}
\end{eqnarray}
where the trace is over the arguments of the shifted auxiliary fields $\phi _i$ with a periodic boundary condition along the imaginary time axis.
Using (\ref{Imeson}) and (\ref{Lmeson}) we find that the mesonic pressure is given by 
\begin{eqnarray}
p_M (T, A_4) 
& = & 
- \frac{T}{2} \sum_n \int \frac{d^3 q}{(2\pi)^3} \left\{ 
\ln \left[ \beta^2 \left( (2 G)^{-1} -  \Pi_\sigma (\omega_n, q, A_4) \right) \right]  \right. \nonumber \\
& & \qquad \qquad \qquad  \qquad 
\left. + 3 \ln \left[ \beta^2 \left( (2 G)^{-1}  -  \Pi_\pi (\omega_n, q, A_4) \right) \right]  \right\}
%\left. \frac{\delta^2 I}{\delta \phi _i \delta \phi _j } \right|_{\phi = \phi _0} 
\label{pressurem}
\end{eqnarray}
%The coefficients of the quadratic term are given by
%\begin{eqnarray}
% {\rm Tr} \ln  
%\left.  \frac{\delta^2 I}{\delta \phi _i \delta \phi _j } \right|_{\phi = \phi _0}  = %V \sum_n \int \frac{d^3 q}{(2\pi))^3}
%\left\{
%\begin{array}{lcl}
%  &    (2G)^{-1} - \Pi_\sigma (\omega_n, q, A_4) \qquad \qquad & \hbox{  for } i = j = 0 \\
 % &    (2G)^{-1}- \Pi_\pi (\omega_n, q, A_4 )     \qquad \qquad & \hbox{ for } i = j= 1, 2, 3 \\
 % &    0    \qquad \qquad & \hbox{ for } i \neq j 
%\end{array}
%\right.
%\end{eqnarray}
where $\omega_n = 2n \pi T$ is the bosonic Matsubara frequency and 
\begin{eqnarray}
\Pi_\sigma (\omega_n, q,  A_4) & = & \Pi_{\sigma}^1 ( A_4 ) + \Pi_{\sigma}^2  (\omega_n, q,  A_4) \\
\Pi_\pi (\omega_n, q,  A_4 ) & = & \Pi_\pi^1 ( A_4)  + \Pi_\pi^2  (\omega_n, q, A_4 ) 
\end{eqnarray}
are the meson self-energies with
\begin{equation}
\Pi_\sigma^1 =  \Pi_\pi^1 = %\Pi_{\sigma}^1 + \Pi_{\sigma}^2 \\
- 2 T \sum_n  \tr_c \psekibun \frac{1}{(\epsilon_n + gA_4)^2 + E_p^2 }  
\end{equation}
and
\begin{eqnarray}
\Pi_\sigma^2  (\omega_n, q,  A_4) & = & \left( \omega_n^2 + \bq^2 + 4M_0^2 \right) F (\omega_n, \bq,  A_4 )
%2 T \sum_{n'} \tr_c \psekibun  
%\frac{\omega_n^2 + \bq^2 + 4M_0^2}{\left[(\epsilon_{n'} + gA_4)^2 + E_p^2\right] \cdot \left[(\epsilon_{n'} + gA_4+ \omega_n)^2 + E_{p+q}^2\right]}  
\\
\Pi_\pi^2 (\omega_n, q,  A_4) & = &  \left( \omega_n^2 + \bq^2 \right) F (\omega_n, \bq,  A_4 )
\end{eqnarray}
where we have defined a dimensionless common multiplicative factor by 
\begin{equation}
F (\omega_n, \bq ,  A_4) = 2 T \sum_{n'} \tr_c \psekibun  
\frac{1}{\left[(\epsilon_{n'} + gA_4)^2 + E_p^2\right] \cdot \left[(\epsilon_{n'} + gA_4+ \omega_n)^2 + E_{p+q}^2\right]} 
\label{F}
\end{equation}

We now show that the contributions of the mesonic correlation to the free energy (\ref{Momega}) or the pressure (\ref{Mpressure}) indeed contain
those of free meson gas composed of massless pions and massive sigma mesons.  
For this purpose we first eliminate the "tadpole" terms $\Pi_\sigma^1$ and $\Pi_\pi^1$ by using the stationary condition, or the "gap equation" (\ref{gap})
which determines the quark quasiparticle mass $M_0$ in the symmetry broken phase.
We then find, 
\begin{eqnarray}
%\left.  \frac{\delta^2 I}{\delta \phi _i \delta \phi _j } \right|_{\phi = \phi _0}  =
%\left\{
%\begin{array}{lcl}
 % &    ( \omega_n^2 + \bq^2 + 4 M_0^2) F(\omega_n, q, A_4) + {\displaystyle \frac{m_0}{2G M_0} } \quad & \hbox{  for } i = j = 0 \\
 % &    ( \omega_n^2 + \bq^2) F(\omega_n, q, A_4)  + { \displaystyle \frac{m_0}{2G M_0} } \quad  & \hbox{ for } i = j= 1, 2, 3 \\
 % &    0    \qquad \qquad & \hbox{ for } i \neq j 
%\end{array}
%\right.
(2 G)^{-1} -  \Pi_\sigma (\omega_n, q, A_4) 
& = &   ( \omega_n^2 + \bq^2 + 4 M_0^2) F(\omega_n, q, A_4) + {\displaystyle \frac{m_0}{2G M_0} } 
\label{smeson}\\
(2 G)^{-1} -  \Pi_\pi (\omega_n, q, A_4) 
& = &   ( \omega_n^2 + \bq^2 ) F(\omega_n, q, A_4) + {\displaystyle \frac{m_0}{2G M_0} } .
\label{pmeson}
\end{eqnarray}

\subsection{Chiral limit: $m_0 = 0$}
We first consider the chiral limit, $m_0= 0$.  
In this limiting case, 
\begin{eqnarray}
%\left.  \frac{\delta^2 I}{\delta \phi _i \delta \phi _j } \right|_{\phi = \phi _0}  =
%\left\{
%\begin{array}{lcl}
 % &    ( \omega_n^2 + \bq^2 + 4 M_0^2) F(\omega_n, q, A_4) + {\displaystyle \frac{m_0}{2G M_0} } \quad & \hbox{  for } i = j = 0 \\
 % &    ( \omega_n^2 + \bq^2) F(\omega_n, q, A_4)  + { \displaystyle \frac{m_0}{2G M_0} } \quad  & \hbox{ for } i = j= 1, 2, 3 \\
 % &    0    \qquad \qquad & \hbox{ for } i \neq j 
%\end{array}
%\right.
(2 G)^{-1} -  \Pi_\sigma (\omega_n, q, A_4) 
& = &   ( \omega_n^2 + \bq^2 + 4 M_0^2) F(\omega_n, q, A_4) 
\label{smeson0}\\
(2 G)^{-1} -  \Pi_\pi (\omega_n, q, A_4) 
& = &   ( \omega_n^2 + \bq^2 ) F(\omega_n, q, A_4) 
\label{pmeson0}
\end{eqnarray}
so that we can separate the contributions of the collective bare meson 
modes from non-collective individual excitations: 
\begin{equation}
p_M (T, A_4) % =  {\rm Tr} \ln  \left.  \frac{\delta^2 I}{\delta \phi _i \delta \phi _j } \right|_{\phi = \phi _0}  
= - \frac{1}{2} T \sum_n \int \frac{d^3 q}{(2\pi)^3}
\left[ \ln \left( \beta^2 ( \omega_n^2 + \bq^2 + 4 M_0^2 ) \right) + 
3 \ln  \left( \beta^2 ( \omega_n^2 + \bq^2) \right)  + 4 \ln F(\omega_n, q, A_4) 
\right]
\end{equation}
The first two terms contain contributions to the pressures, respectively, of free sigma meson gas with mass $m_\sigma = 2 M_0$ and of free massless pion gas with 3 isospin degeneracy.  
Indeed, one can perform each bosonic Matsubara frequency sum by contour integration 
%as we did for fermionic Matsubara frequency sum of quarks, 
and find, 
\begin{eqnarray}
p_M^{\rm free} (T) 
%& \equiv & - \frac{1}{2} T \sum_n \int \frac{d^3 q}{(2\pi)^3}
%\left[ \ln  \left( \beta^2 ( \omega_n^2 + \bq^2 + 4 M_0^2 ) \right) 
%+ 3 \ln \left( \beta^2  ( \omega_n^2 + \bq^2) \right) \right]  \nonumber \\
&  =  &
p_M^0 + \int \frac{d^3 q}{(2\pi)^3}
\left[ \frac{q^2}{3\omega_q} f_B (\omega_q) + 3 \times \frac{q}{3}  f_B (q) \right]
\end{eqnarray}
where $\omega_q = \sqrt{\bq^2 + 4 M_0^2}$ is the sigma meson energy and $f_B (\omega) = 1/( e^{\beta \omega} -1 )$ is a bosonic single particle distribution function.   
The first term is the mesonic vacuum pressure $p_M^0$,
\begin{equation}
p_M^0 = - \frac{1}{2} \int^{\Lambda_b} \frac{d^3 q}{(2\pi)^3}\left[ 3 q + \omega_q \right]
\end{equation}
where we have indicated an additional cutoff $\Lambda_b$ which need to be introduced to regularize otherwise-divergent mesonic momentum integral. 
This vacuum mesonic pressure needs to be removed by renormalization.    
The momentum integral for the pressure of the massless pion gas can be evaluated analytically and one finds a familiar Stephan-Boltzmann pressure  
with three-fold isospin degeracy:
\begin{equation}
p_{\rm pion}^{\rm free} = \int \frac{d^3 q}{(2\pi)^3} 3 \times \frac{q}{3}  f_B (q) = 3 \times \frac{\pi^2}{90} T^4 
\label{free_pion}
\end{equation}

Now we evaluate the correction to the free meson gas due to underlying quark substructure of mesons.
This is contained in the additional contribution, 
\begin{equation}
\Delta p_M (T, A_4) = - 2 T \sum_n \int \frac{d^3 q}{(2\pi)^3} \ln F(\omega_n, q, A_4) 
\label{noncollective}
\end{equation}
where  $\omega _n$ is the bosonic Matsuibara frequencies, $\omega _n=2\pi nT$.
The function $F$ defined by (\ref{F}) also contains the sum over the fermionic Matsubara frequencies $\epsilon_n = (2n +1) \pi T$.
These sums are computed by the contour integration on a complex $z$ plane with multiplicative complex function 
$\pm 1/(\exp (z \beta) \pm 1)$ which has poles on the imaginary $z$-axis at $\epsilon_n$ ($\omega_n$). 
We find
\begin{eqnarray}
\Delta p_M (T, A_4) 
& = & - 2 \int_{C} \frac{d z}{2\pi i} \int \frac{d^3 q}{(2\pi)^3} \frac{1}{e^{z\beta}-1} \ln {\cal F} (z, q, A_4) .
\label{}
\end{eqnarray}
where
\begin{equation}
{\cal F} (z, q, A_4) = F (- i z, q, A_4)
\end{equation}
Noting that the function $\cal F$ has poles only on the real $z$axis, we modify the integration path continuously to the one which encircles the real $z$ axis clock-wise.  
Rewriting the real value of $z$ by $\omega$,  we obtain
\begin{eqnarray}
\Delta p_M (T, A_4) 
& = &- 2 \int_{-\infty}^\infty  \frac{d \omega }{2\pi i} \int \frac{d^3 q}{(2\pi)^3} \frac{1}{e^{\omega\beta}-1}  
\ln \frac{{\cal F} ( \omega+i \epsilon, q, A_4)}{{\cal F} (\omega - i \epsilon, q A_4 ) } 
\nonumber \\
& = & - 2 \int_0^\infty \frac{d z}{2\pi i}  \int \frac{d^3 q}{(2\pi)^3} (1 + \frac{2}{e^{\omega\beta}-1} ) \ln 
\frac{{\cal F} ( \omega+i \epsilon, q, A_4)}{{\cal F} (\omega - i \epsilon, q A_4 ) } ,
\end{eqnarray}
where we have used the fact that the function ${\cal F} (\omega, q , A_4)$ is a even function of $\omega$ and 
$ 1/ (e^{-\omega \beta} -1) = 1+ 1/ (e^{\omega \beta} -1) $. 

Similarly, the sum over the fermionic Matsubara frequency in (\ref{F}) can be performed by contour integration:
\begin{eqnarray}
{\cal F} ( \omega, q, A_4) & = &   \tr_c \psekibun  \int _{\mathcal{C}} \frac{dz}{2\pi i}  \frac{1}{e^{z\beta} + 1 } 
\frac{1}{\left[  ( - (z + i gA_4)^2 + E_p^2\right] \cdot \left[ - (z + i gA_4 + \omega)^2 + E_{p+q}^2\right]}  
\nonumber \\
& = &   \tr_c \psekibun  \frac{1}{2E_p 2E_{p+q}}
 \frac{1}{e^{(E_p -iA_4) \beta} + 1 } 
\left( \frac{1}{\omega+ E_p - E_{p +q}} - \frac{1}{\omega - E_p + E_{p +q}} \right) 
+  ... 
\nonumber 
\end{eqnarray}
where we have performed the contour integration by modifying the path to the paths encircling the poles on the real $z$ axis, showing only one of four terms.
Again, the external gauge fields appear as a phase factor in the quark distribution function in 
${\cal F} (\omega _n, q, A_4)$ as in the mean-field approximation.  
We therefore replace these phase factors by the Polyakov loops and then substitute them by statistical average 
as in (\ref{averagedis}) performing sum over bosonic Matsubara frequencies by contour integration, we found  
\begin{eqnarray}
{\cal F} ( \omega, q ) & \equiv & \langle {\cal F} ( \omega, q,  A_4 ) \rangle  
= {\cal F}_{\rm scat} (\omega, q) + {\cal F}_{\rm pair} (\omega, q ) 
\label{calF}
\end{eqnarray}
where  
 \begin{eqnarray}
 {\cal F}_{\rm scat} ( \omega, q  ) 
%& \equiv & \langle F^{\rm scat} (q, \omega+ i \epsilon, A_4  ) \rangle  \nonumber \\
& = & 3 \int \frac{d ^3 p}{(2\pi)^3} \frac{1}{2 E_p 2 E_{p+q}} 
\left( \frac{1}{\omega+ E_p - E_{p +q}} - \frac{1}{\omega - E_p + E_{p +q}} \right) 
\nonumber \\
& & \qquad \qquad \times  \left( f_\Phi  (E_p) - f_\Phi (E_{p+q}) \right)  \\
{\cal F}_{\rm pair} (\omega, q ) 
%& \equiv  & \langle F^{\rm pair} (q, \omega+ i \epsilon, A_4  ) \rangle \nonumber \\
& = & 3 \int \frac{d ^3 p}{(2\pi)^3} \frac{1}{2 E_p 2 E_{p+q}} 
\left( \frac{1}{\omega + E_p + E_{p +q}} - \frac{1}{\omega - E_p - E_{p +q}} \right) 
\nonumber \\
& & \qquad \qquad \times  \left( 1 - f_\Phi (E_p) - f_\Phi (E_{p+q} ) \right)  .
\end{eqnarray}
The detail of this computation is given in the appendix A.
We may interpret these correlation energy as non-collective fluctuations of the system carrying 
mesonic quantum numbers with quenched quark distributions. 

For the computation of the correlation energy or pressure, it is convenient to decompose the function 
${\cal F} (\omega \pm i \epsilon, q ) $ into real part ${\cal F}_1( \omega, q) $ and imaginary part 
$ {\cal F}_2 ( \omega, q ) $: 
\begin{eqnarray}
{\cal F} ( \omega \pm i \epsilon, q ) & = & {\cal F}_1( \omega, q) \pm i {\cal F}_2 (\omega, q ) 
%\nonumber \\
%{\cal F} ( \omega \pm i \epsilon, q ) =| {\cal F} (q, \omega \pm i \epsilon ) | e^{\pm i \phi (q, \omega) } 
=  \sqrt{  {\cal F}_1( \omega, q )^2 +  {\cal F}_2 ( \omega, q )^2 } e^{\pm i \phi ( \omega, q) }  ,
\end{eqnarray}
%with the modulus and the argument given by
%\begin{eqnarray}
%& & | {\cal F} (q, \omega \pm i \epsilon ) | = \sqrt{  {\cal F}^1(q, \omega)^2 +  {\cal F}^2 (q, \omega )^2 } \\
%\end{eqnarray}
where the argument $\phi$ is given by
\begin{equation}
 \phi (\omega, q) =  \tan^{-1} \frac{{\cal F}_2(\omega, q)}{{\cal F}_1(\omega, q)} ~~. \label{eq:phi}
\end{equation}

The pressure arising from the non-collective or individual excitations of the system is given by 
\begin{equation}
\Delta p_{M} (T) = 
\langle \Delta p_M (T, A_4) \rangle % 4 T \int \frac{d^3 q}{(2\pi)^3} \ln F(\omega_n, q, A_4) 
 =  - 2 \int^{\Lambda_b} \frac{d ^3 q}{(2\pi)^3} \int_0^\infty \frac{d \omega}{2\pi}
\left[ 1 + \frac{2}{ e^{\beta \omega} -1} \right] 2 \phi (\omega, q) .
\label{eq:phase}
%\ln \left[ \frac{F(q, \omega + i \epsilon, A_4)}{F(q, \omega- i \epsilon, A_4)} \right]
\end{equation}

\subsection{Breaking the chiral symmetry with $m_0 \ne 0$}

Foregoing analysis applies only to the chiral limit ($m_0 = 0$) and symmetry broken phase.
For non-zero value of $m_0$, the separation of the collective modes and the individual excitations is not 
as simple as above analysis due to the term $m_0/(2G M_0)$ in the dispersions (\ref{smeson}) 
and (\ref{pmeson}) for each kind of mesons. 
We expect in this case that the pions, would-be Nambu-Goldstone modes associated with spontaneous symmetry breaking of the chiral symmetry, acquire non-zero mass, $m_\pi$. 
To calculate the contribution to the thermodynamic potential of such modes we need to go back to the original formula 
(\ref{pressurem}) for the pressure from mesonic correlation. 
It is convenient to write 
\begin{eqnarray}
M_{\sigma} (\omega_n, q ) & = & (2 G)^{-1}  - \langle  \Pi_{\sigma} (\omega_n, q, A_4) \rangle \\
M_{\pi} (\omega_n, q ) & = & (2 G)^{-1}  -  \langle \Pi_{\pi} (\omega_n, q, A_4) \rangle 
\end{eqnarray}
or using the gap equation, 
 \begin{eqnarray}
M_{\sigma} (\omega_n, q) & = &  ( \omega_n^2 + \bq^2 + 4 M_0^2) \langle F (\omega_n, q, A_4) \rangle + {\displaystyle \frac{m_0}{2G M_0} } \\
M_{\pi} (\omega_n, q) & = &  ( \omega_n^2 + \bq^2) \langle F (\omega_n, q, A_4) \rangle 
+ {\displaystyle \frac{m_0}{2G M_0} } 
\label{M_sigma}
\end{eqnarray}
Performing the bosonic-Matsubara frequency sum by contour integration again, we obtain
\begin{eqnarray}
p_M (T) % 4 T \int \frac{d^3 q}{(2\pi)^3} \ln F(\omega_n, q, A_4) 
 & = & -  \int \frac{d ^3 q}{(2\pi)^3} \frac{1}{2\pi i}\int_0^\infty d \omega
\left[ 1 + \frac{2}{ e^{\beta \omega} -1} \right]  \nonumber \\ 
& & \qquad \qquad \times \left\{ 
3~ \ln \left[ \frac{{\cal M}_\pi( \omega + i \epsilon, q)}{{\cal M}_\pi ( \omega- i \epsilon, q)} \right]  
+ \ln  \left[  \frac{{\cal M}_\sigma( \omega + i \epsilon, q)}{{\cal M}_\sigma( \omega- i \epsilon, q)} \right] \right\}
\label{Pmm}
\end{eqnarray}
where we have written
\begin{eqnarray}
\mathcal{M}_{\pi/\sigma} (\omega , q) = M_{\pi/\sigma} ( - i\omega _n, q) .
\end{eqnarray}
with $\langle F (\omega_n, q, A_4) \rangle$ in $M_{\pi/\sigma} ( - i\omega _n, q)$ replaced by (\ref{calF}). 
Expressing the integrand of (\ref{Pmm}) in terms of the arguments %$\phi _{\pi/\sigma} (\omega , q)$ 
of complex $ {\cal M}_{\pi/\sigma} ( \omega \pm i \epsilon, q) $, we have
%$\mathcal{M}_{\pi/\sigma} (\omega + i\epsilon , q)$: 
%%%%%%%%%%%%%%%%%%%%%%%%%%%%%%%%%%%%%%%%%%%%%%%%%%
\begin{eqnarray}
p_M(T) = - \int ^{\Lambda _b}\frac{d^3q}{(2\pi )^3} 
 \int_0^\infty \frac{d\omega }{2\pi }\Bigl[ 1+\frac{2}{e^{\omega\beta}-1}\Bigr]  
\left[ 3 \phi_\pi (\omega , q) +\phi_\sigma (\omega, q) \right].
\end{eqnarray}
where
\begin{eqnarray}
\phi _{\pi/\sigma} (\omega , q)
& = & \mbox{tan}^{-1}\frac{ \mathcal{M}_{\pi/\sigma}^2 (\omega , q)} { \mathcal{M}_{\pi/\sigma}^1(\omega , q)} 
%\phi _{\sigma }(\omega , q) 
%& = & \mbox{tan}^{-1}\frac{\mathcal{M}_{\sigma}^2(\omega , q)}{\mathcal{M}_{\sigma}^1(\omega , q)} 
\label{phipi_phisigma}
\end{eqnarray}
with
\begin{eqnarray}
{\cal M}_{\pi/\sigma} ( \omega \pm i \epsilon, q) 
& = & {\cal M}^1_{\pi/\sigma} ( \omega, q) \pm i {\cal M}^2_{\pi/\sigma} ( \omega, q) . \label{phi_of_M}%\\
%  {\cal M}_\sigma ( \omega \pm i \epsilon, q) & = & {\cal M}^1_\sigma ( \omega, q) \pm i  {\cal M}^2_\sigma ( \omega, q)
\end{eqnarray}

We note that unlike the case of the chiral limit, 
$\phi _{\pi }$ and $\phi _{\sigma }$ defined as eq.(\ref{phipi_phisigma}) contain contributions not only 
from non-collective modes but also from collective modes. 
Here we call the collective meson mode only when it appears as an isolated pole on the real $\omega$ axis: 
it requires the condition that both real and imaginary parts of  $\mathcal{M}_{\pi/\sigma} (\omega , q)$ vanish 
for a certain combination of $\omega$ and $q$ determining the mesonic dispersion relation. 
In this case the argument $\phi (\omega, q)$ becomes a $\delta$ function, $\delta ( \omega - \omega_q) $ 
with the mesonic dispersion relation, $\omega_q \simeq \sqrt{q^2 + m^2}$, with $m = m_\pi (m_\sigma) $.

On the other hand, if only the real part of  ${\cal M} (\omega, q) $ vanishes while the imaginary part is non-vanishing, such mode decays by Landau-damping into non-collective excitations.     
We refer these modes as the non-collective individual excitations  
%More detail of the computation is given in appendix C.
%subsection{Mesonic correlations in the high temperature phase}
As the temperature increases the low energy boundary of the continuum of the particle-antiparticle 
excitations comes down and eventually absorbs the meson poles.    
Hence no isolated meson poles remain. 
In the chiral limit  this happens at $T=T_c$ where the chiral symmetry is restored and pions,  
Nambu-Goldstone modes, disappear. 
In this case, there is no meson pole contribution and only contribution from cuts remains in the mesonic correlation energy.  
With explicit chiral symmetry breaking by the bare quark mass, we have to go back again to the original expression (\ref{pressurem}) in order to compute the 
mesonic correlation energy as has been done in \cite{RHRW08}. 
We found numerically that the real part of  ${\cal M} (\omega, q) $ does not vanish at high 
temperatures for all values of $(\omega,q)$, showing no sign of persistence of mesonic modes,
 as shown in the section 6.

\subsection{Cut-off parameters} \label{sec:cutoff}

Since our model contains some divergences, we have to introduce cut-off parameters to suppress these divergences. For calculating pressure of mesonic correlation, we need two cut-off parameters.
First divergence appears under the mean field approximation.
A cut-off parameter is introduced in the eq.(\ref{condpress}). This is the vacuum pressure under the mean field approximation. The first term of eq.(\ref{condpress}) diverges because of the momentum integral. We remove higher momentum modes by the cut-off $\Lambda_f$. 
The thermal excitation term, the third term on the right hand side of
eq.(\ref{mfpressure}), doesn't need any cut-off because quark distribution functions suppress the excitations of high momentum modes. Hence, when once $\Lambda _f$ is introduced in the vacuum term, we can calculate the pressure under the mean field approximation without any more divergence. 

However, when we calculate pressure of the mesonic correlation, additional divergence appears 
in the  momentum integral with the first term in the bracket  of eq.(\ref{eq:phase}). This divergence arises from the meson momentum integral and we cannot eliminate it by quark momentum cut-off $\Lambda _f$ alone. This problem arises because  NJL type models are unrenormalize: even if we suppress a divergence from a calculation of some order, additional divergence appears in the next order calculation. 
For this reason,  we have to introduce second cut-off parameter $\Lambda _b$ to suppress this divergence from bosonic loop calculation. 
Since $\Lambda _b$ has no relation to $\Lambda _f$ mathematically, we should determine the value of it based on physical considerations. In this work, we choose $\Lambda _b=2\Lambda _f$ \cite{FB96}. 

The integral with the second term in the bracket of eq.(\ref{eq:phase}) doesn't diverge. Since bosonic distribution function as a function of $\omega $ suppresses excitation modes at large $\omega $ region, no divergence appears in the $\omega $ integral. As for $q$ integral, it seems at first sight to diverge because the bosonic distribution function is only a function of $\omega $. However since $\phi $ becomes very small at large $q$, the divergence doesn't  appear. One can see this from eq.(\ref{N_scat}). When we fix $\omega $ and see large $q$ region, it is always space-like region so that the behavior of $\mathcal{F}_2 $ corresponds with $\mathcal{F}_{scatt, 2}$.  From eq.(\ref{N_scat}), we see that $N_{scat}(\mathcal{\epsilon })$ becomes exponentially small at large $q $ region. As a result $\mathcal{F}_{scat, 2}$ also gets close to zero from eq.(\ref{F_scat2}). Although eq.(\ref{N_scat}) is an approximation under the condition of small $\epsilon $, this approximation 
 is justified as long as q becomes large for fixed $\omega $ in the $q$ integral of eq.(\ref{eq:phase}).

\section{Numerical results}

\subsection{Pressure}

\begin{figure}[htbp]
\begin{center}
%\resizebox*{!}{4.7cm}{
 \includegraphics[clip,width=55mm, angle=270]{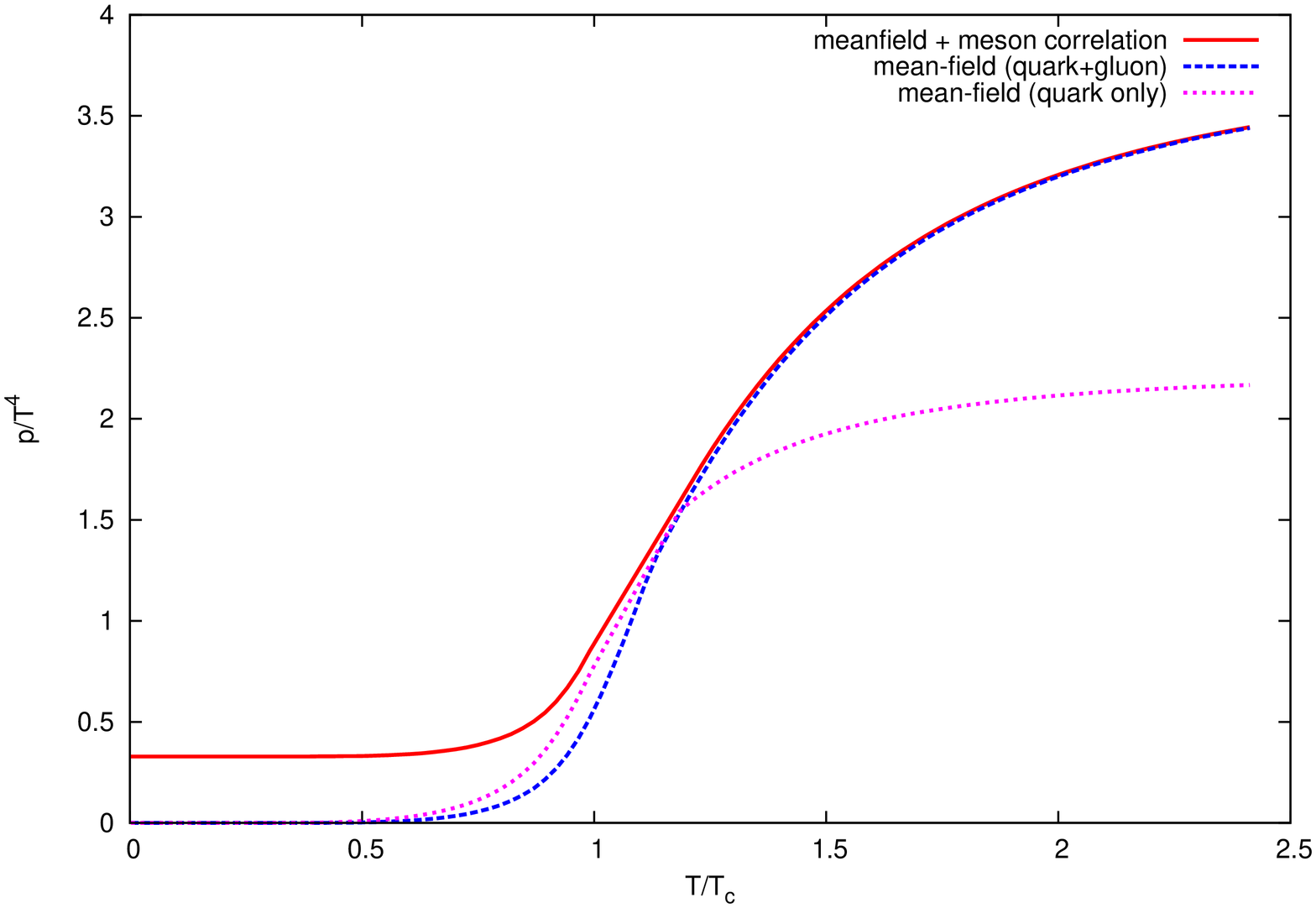}
 \includegraphics[clip,width=55mm, angle=270]{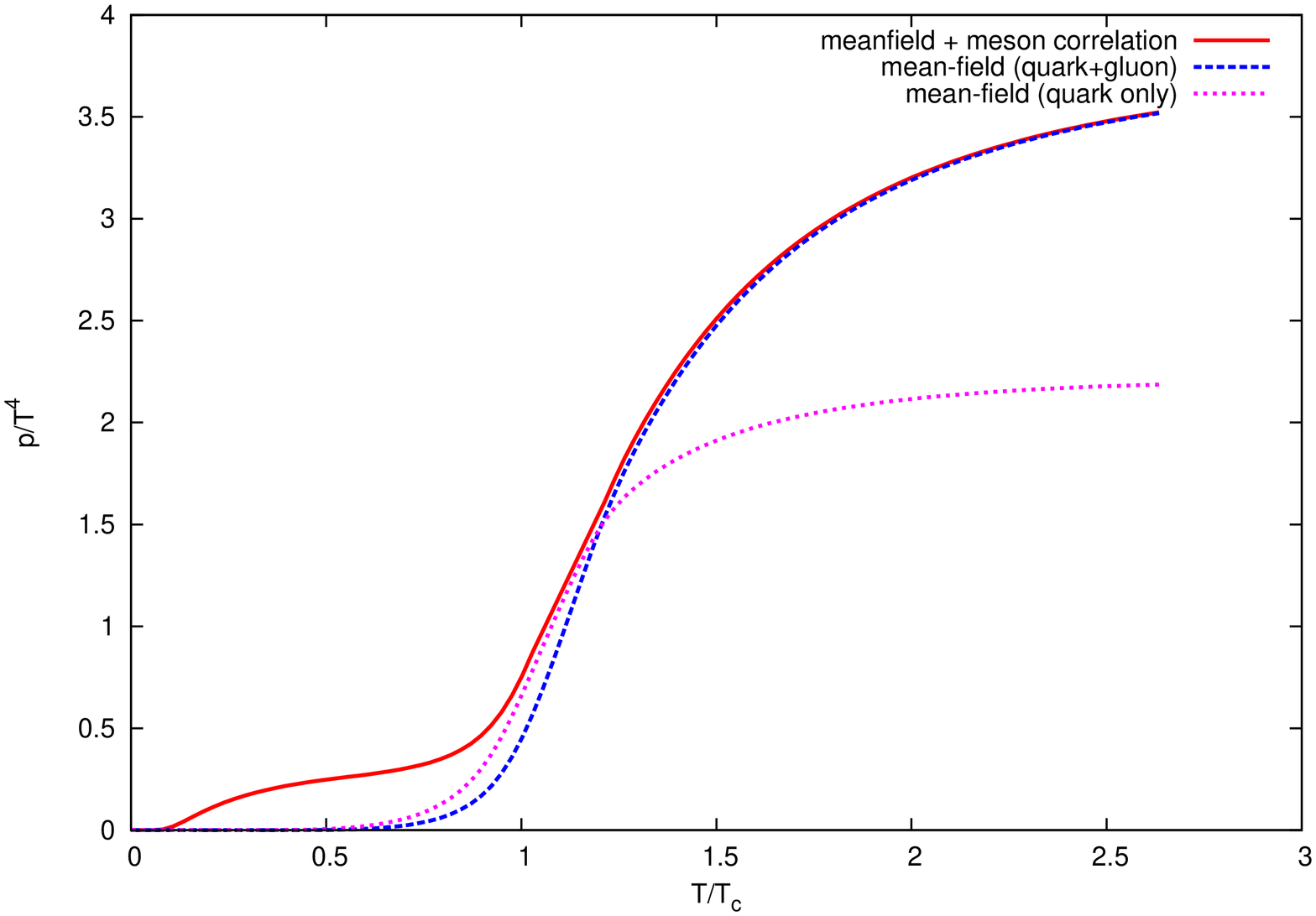}
\end{center}
\caption{
Pressure scaled by $T^4$ as a function of temperature: 
the left panel is for vanishing bare quark mass (the chiral limit) and the right panel is with finite bare quark mass $m_0$ which is chosen to reproduce the pion mass $m_\pi = 140$MeV.  
In comparsison, the pressures calculated in the mean-field approximation are shown with (without) gluon 
contributions. 
}
\label{fig:EOS}
\end{figure}

We show in figure \ref{fig:EOS} the pressure calculated by the present method.  The left panel is the result computed in the 
chiral limit $m_0 = 0$.  At low temperatures below the second order chiral transition temperature $T_c$, 
the pressure from quark quasiparticle excitations are suppressed by the quenching of the distribution function:
only excitations of triad of quarks with color singlet configuration are allowed with effective excitation energy 
three times that of massive quark excitation, this contribution to the pressure is hence suppressed. 
The dominant mode of excitations which determine the pressure is due to massless pion excitation with 
3 isospin degeneracy as shown in the red solid curve.  
We found the contribution from massive sigma meson excitations and non-collective individual excitations of 
the quark triads and antiquark triads are also negligible at low temperature.    
At high temperatures, massless quark excitations in the mean-field approximation dominate the pressure 
with comparable contributions of the gluon pressure phenomenologically introduced in the construction of 
the effective potential. 
Contribution from the mesonic correlations becomes negligible at high temperatures, and the pressure 
approaches to that of ideal gas of massless quarks and gluons in our model.
We note that at large temperatures the cut-off $\Lambda$ in the quark momentum integration reduces the pressure 
compared to that of free quark-gluon gas, as seen in the effective reduction of the Stephan-Boltzmann constant 
$p/T^4$ from $ a = 3 \times 2 \times 2 \times 2 \times 7/8 \times \pi^2/90 = 2.3$.
With symmetry breaking finite bare quark mass, the pion becomes massive and its temperature decrease 
exponentially at very low temperature, otherwise the behavior of pressure at higher temperatures is
qualitatively unchanged.

%\section{Interpretation of numerical results}
\subsection{Collective modes and non-collective modes}
%In the previous subsection, we have shown that mesonic excitations dominate the pressure at low temperatures while they don't contribute to the pressure so much at high temperatures. 
Since mesonic excitations contain two types of modes, collective meson modes and non-collective individual excitations, we consider in this subsection the two contributions separately, for the chiral limit where the separation is manifest, and for the case with the chiral symmetry breaking, where we need to separate them carefully due to the mixing via the bare quark mass term.

\subsubsection{Chiral limit}

As shown in figure 2, the pressure at low temperatures satisfies the Stephan-Boltzmann relation $p/T^4\simeq 0.33$. 
It means that pressure at low temperature comes only from pion collective modes, eq.(\ref{free_pion}), and the contributions from sigma meson collective modes and non-collective modes are very small.  
The contribution from thermal excitations of sigma meson is suppressed compared to that of pions due to large sigma meson mass, $m_\sigma = 2M_0$.
The contribution from non-collective modes is also expected to be suppressed due to the same reason: 
the effective mass of quark triad is $3M_0$ as it  appears in the quenched quark distribution.  
Since the contribution of non-collective modes to the pressure comes from the phase space integral of the argument 
$\phi (\omega, q)$ of  $\mathcal{F} (\omega +i \epsilon, q)$, we like to inspect in detail the behavior 
of $\mathcal{F} (\omega +i \epsilon, q)$.

\begin{figure}[htbp]
\begin{center}
 \includegraphics[clip,width=80mm]{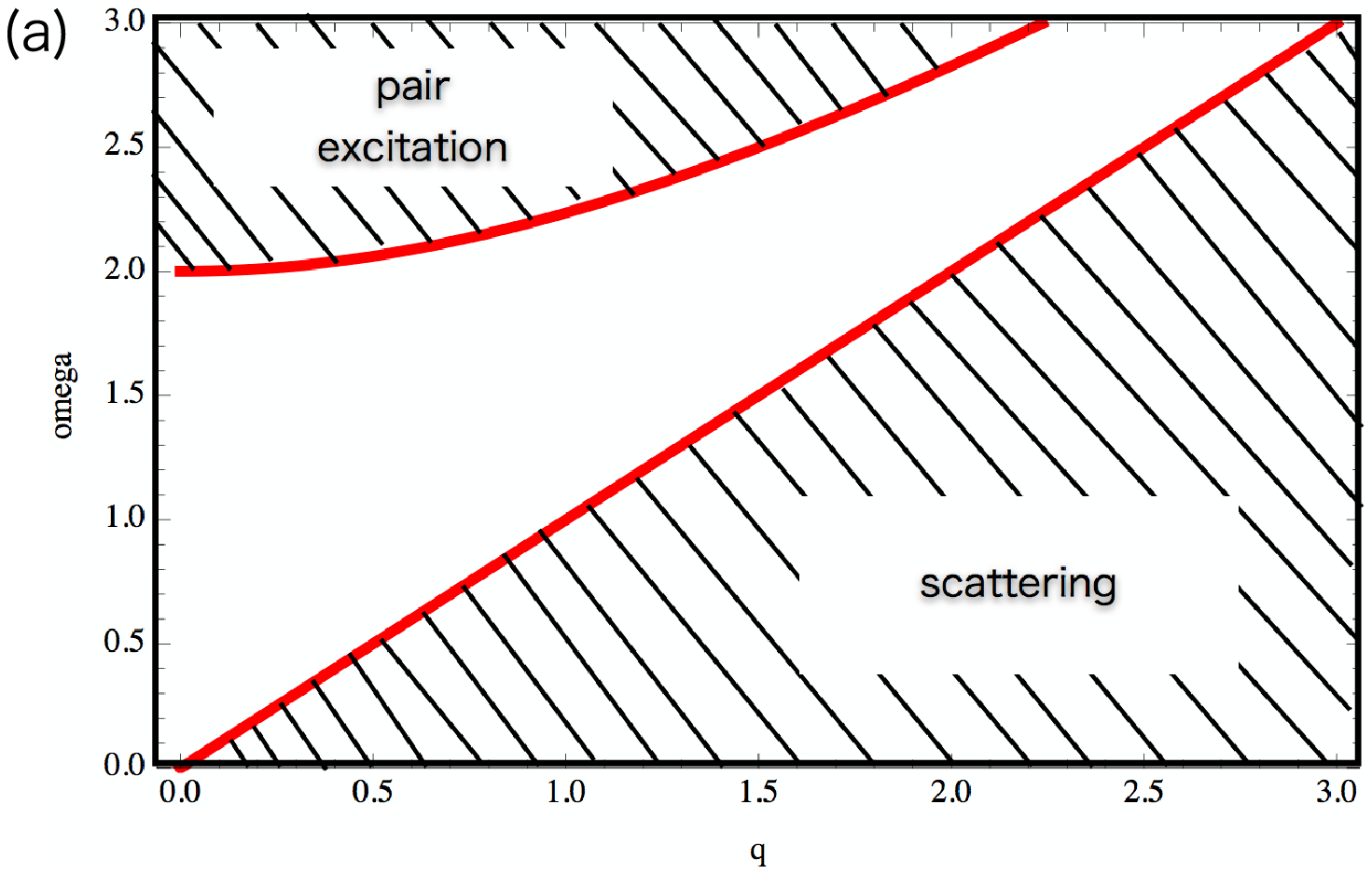}
 \includegraphics[clip,width=80mm]{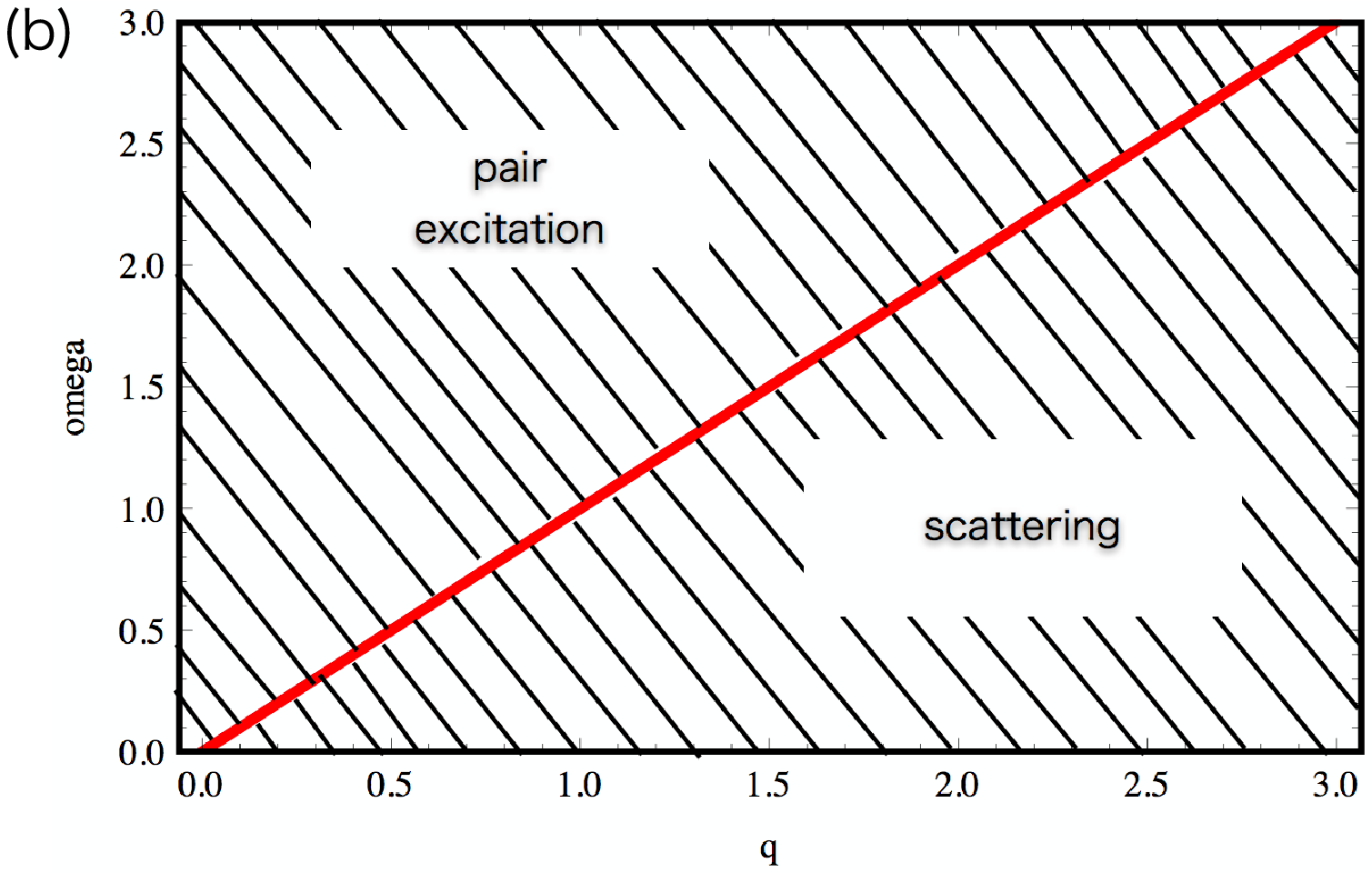}
\end{center}
\caption{Kinematical regions where $\mathcal{F}_2$ is non-vanishing is shown by shaded areas: 
(1) $\omega < q$ (scattering), (2) $\omega > \sqrt{q^2+(2M_0^2)}$ (pair excitation). The left panel is for $T= 0.7 T_c$,  while the right panel is for $T=1.4T_c$. }
\label{fig:F2}
\end{figure}

We show  first the region where the imaginary part of function 
$\mathcal{F}(\omega +i\epsilon  ,q)$, $\mathcal{F}_2$, becomes non-zero in figure \ref{fig:F2}.
It consists of two regions, one in the space-like region $(\omega <q)$ corresponding to the scattering, 
and the other in the time-like region $(\omega> q)$ where pair excitations contributes. 
The upper boundary of scattering region is at the light cone ($\omega = q$) and
the lower boundary of pair excitation region is given by $\omega =\sqrt{q^2+(2M_0)^2}$ where $q$ is the 
momentum of the mesonic excitation and $M_0$ is the constituent quark mass determined by  
the gap equation (\ref{GE}). 
In the region between the two boundaries, $\mathcal{F}_2$ becomes zero.  
%Since the physical meaning of $\mathcal{F}_2$ is number of states of excitations, built upon the thermal quark distributions in the mean-field approximation, with momentum $q$ and energy $\omega $ (Appendix B), 
%it implies that 
There are no physically allowable excitations in the kinematical region between these boundaries, besides meson poles which are not shown here. 
%Since pressure of non-collective modes is given by eqs. (\ref{eq:phi}) and (\ref{eq:phase}), 
%there is no contribution to the pressure from this region.% where $\mathcal{F}_2$ vanishes.
As temperature increases, the upper boundary goes down because increase of temperature causes 
decrease of constituent quark mass $M_0$ as a result of restoration of the chiral symmetry.
Hence there appears individual excitations filling this region at high temperature as shown in the right panel 
of figure \ref{fig:F2}. 
\begin{figure}[htbp]
\begin{center}
%\resizebox*{!}{4.7cm}{
 \includegraphics[clip,width=80mm]{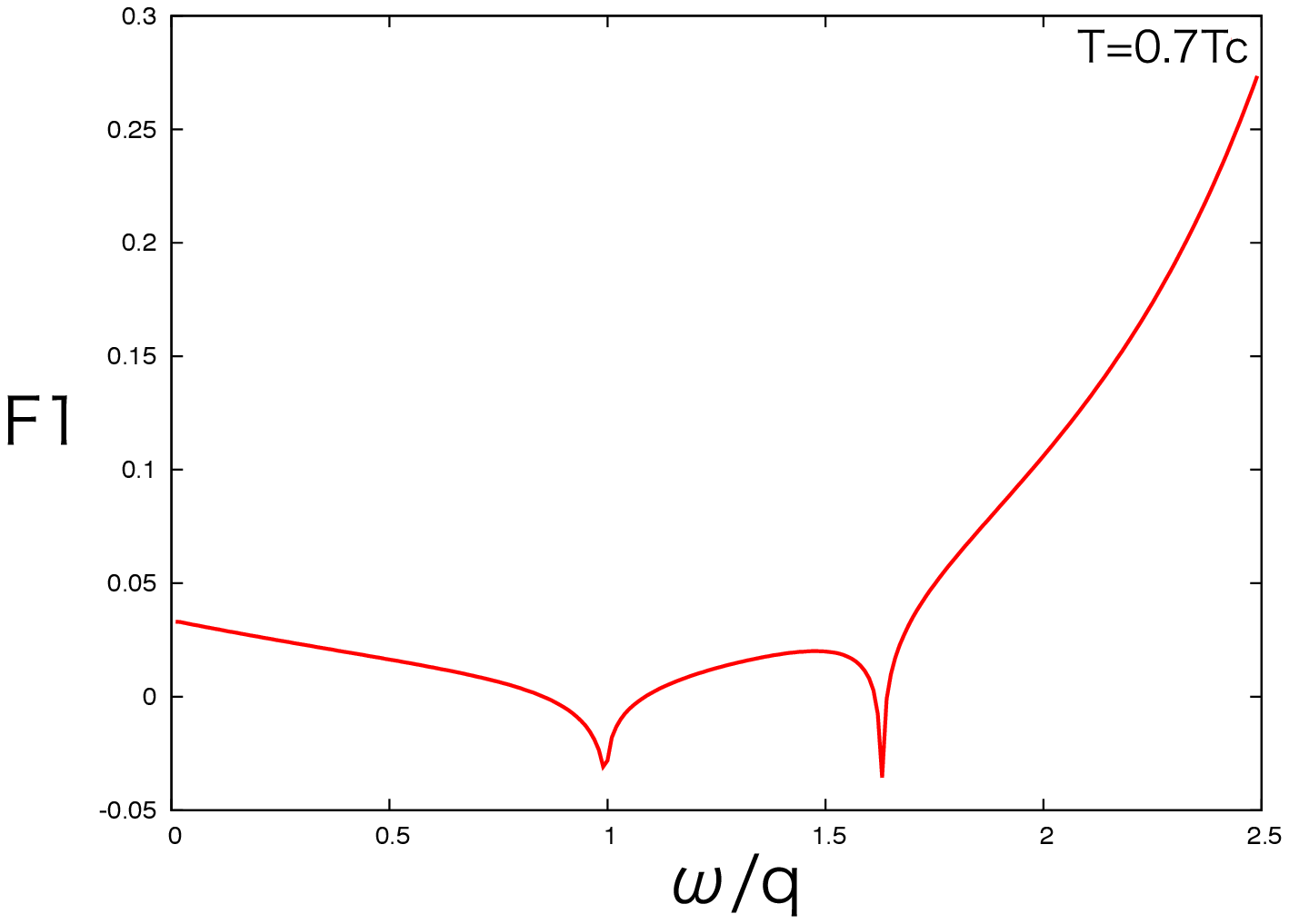}
 \includegraphics[clip,width=80mm]{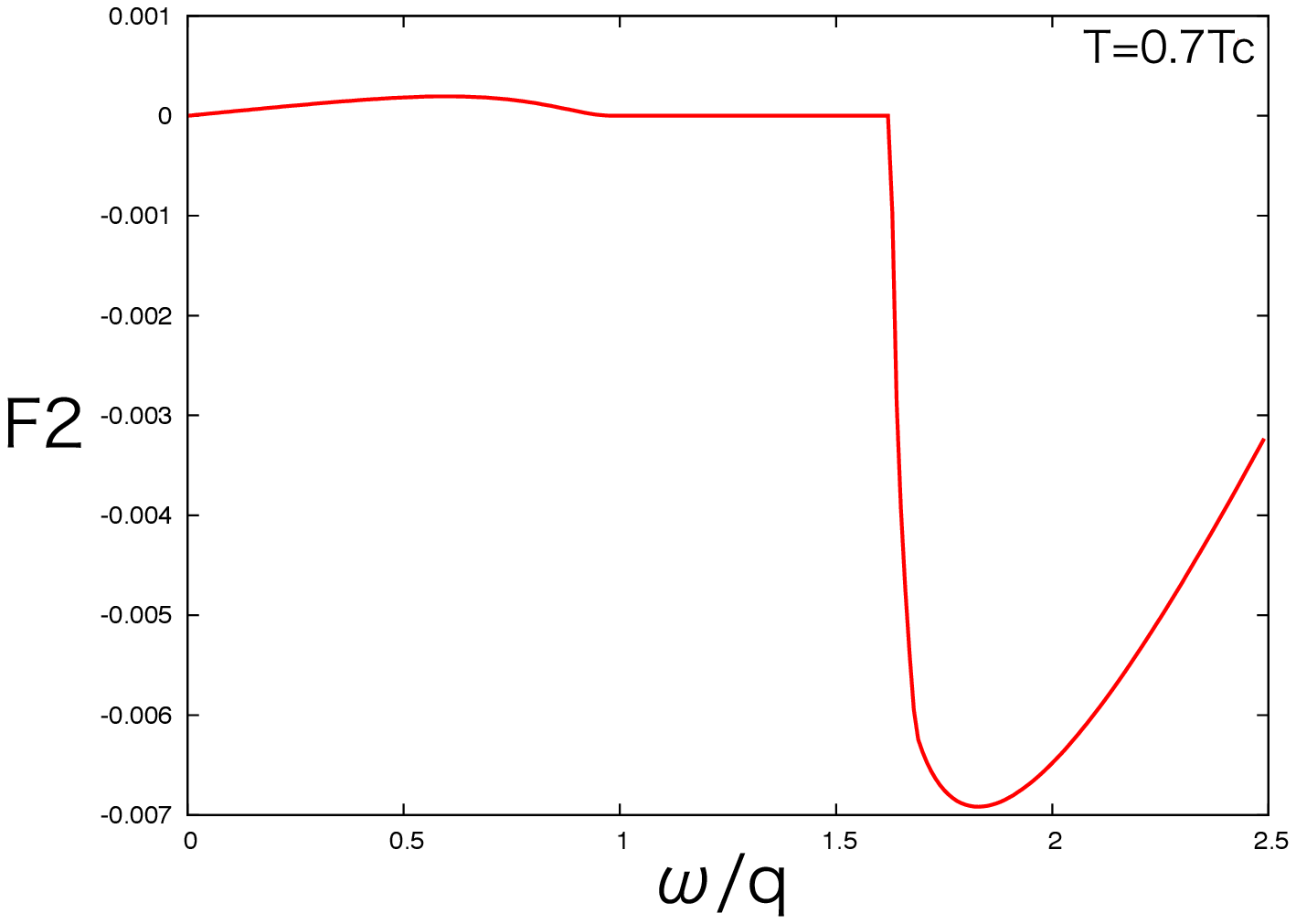}
\end{center}
\caption{Left: $\mathcal{F}_1$ as a function of  $\omega /q$ at $T=0.7T_c$. Right: $\mathcal{F}_2$ as a function of $\omega /q$ at $T=0.7T_c$.}
\label{fig:F1F2LT}
\end{figure}

\begin{figure}[htbp]
\begin{center}
%\resizebox*{!}{4.7cm}{
 \includegraphics[clip,width=80mm]{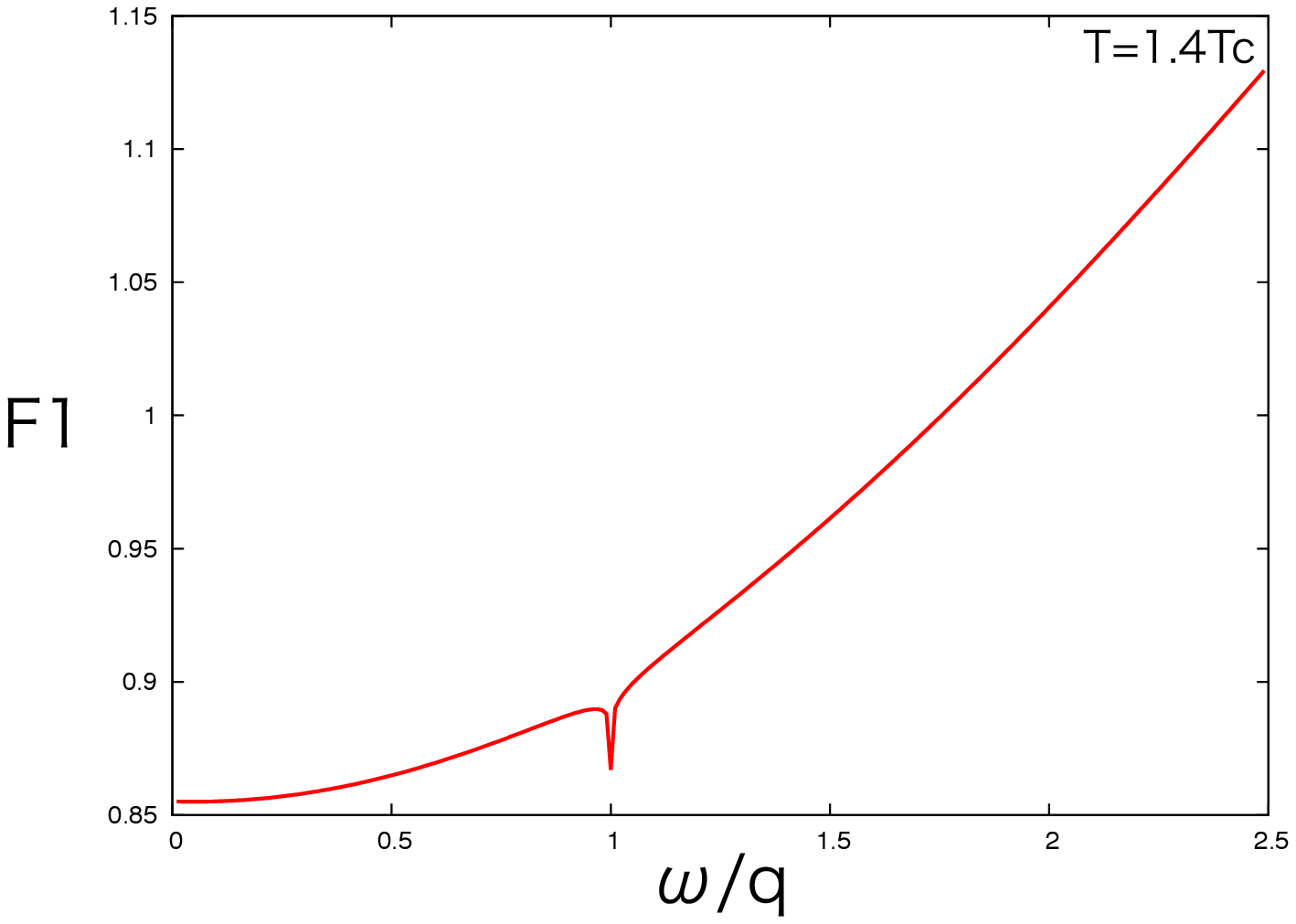}
 \includegraphics[clip,width=80mm]{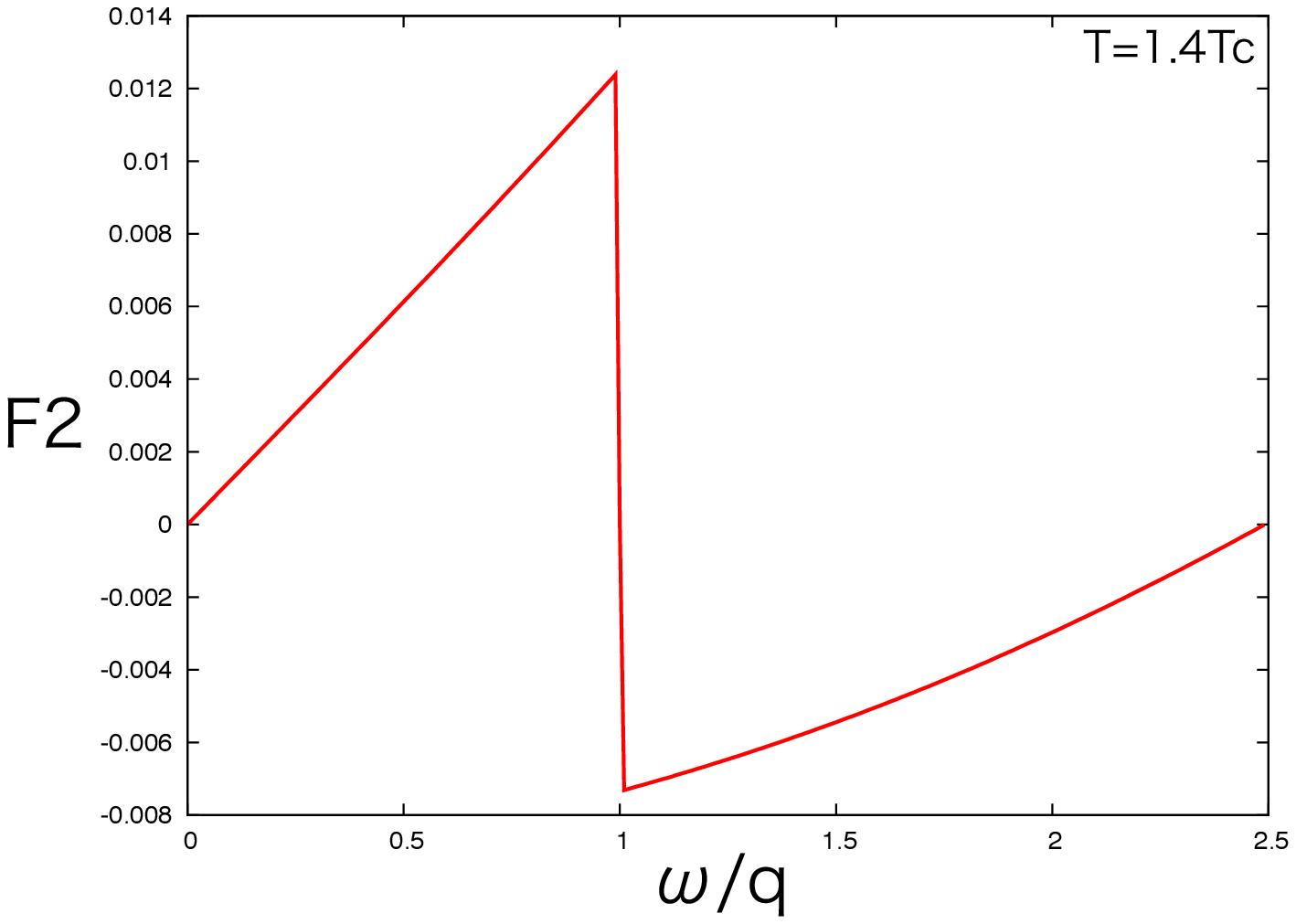}
\end{center}
\caption{Left: $\mathcal{F}_1$ as a function of $\omega /q$ at $T=1.4T_c$. Right: $\mathcal{F}_2$ as a function of $\omega /q$ at $T=1.4T_c$.}
\label{fig:F1F2HT}
\end{figure}

Next we display $\mathcal{F}_1$ and $\mathcal{F}_2$ as a function of $\omega /q$ for a fixed value of $q = \Lambda_f$ at two characteristic temperatures in figure \ref{fig:F1F2LT} ($T=0.7T_c$) and figure \ref{fig:F1F2HT} ($T=1.4T_c$). 
Below $T_c$,  $\mathcal{F}_1$ crosses the horizontal axis and there are two valleys at $\omega /q=1$ and 
$\omega /q\simeq 1.6$, on the other hand, above $T_c$, it doesn't cross the horizontal axis and the valley at $\omega /q\simeq  1.6$ goes down and is absorbed into continuum. 

\begin{figure}[htbp]
\begin{center}
%\resizebox*{!}{4.7cm}{
 \includegraphics[clip,width=110mm]{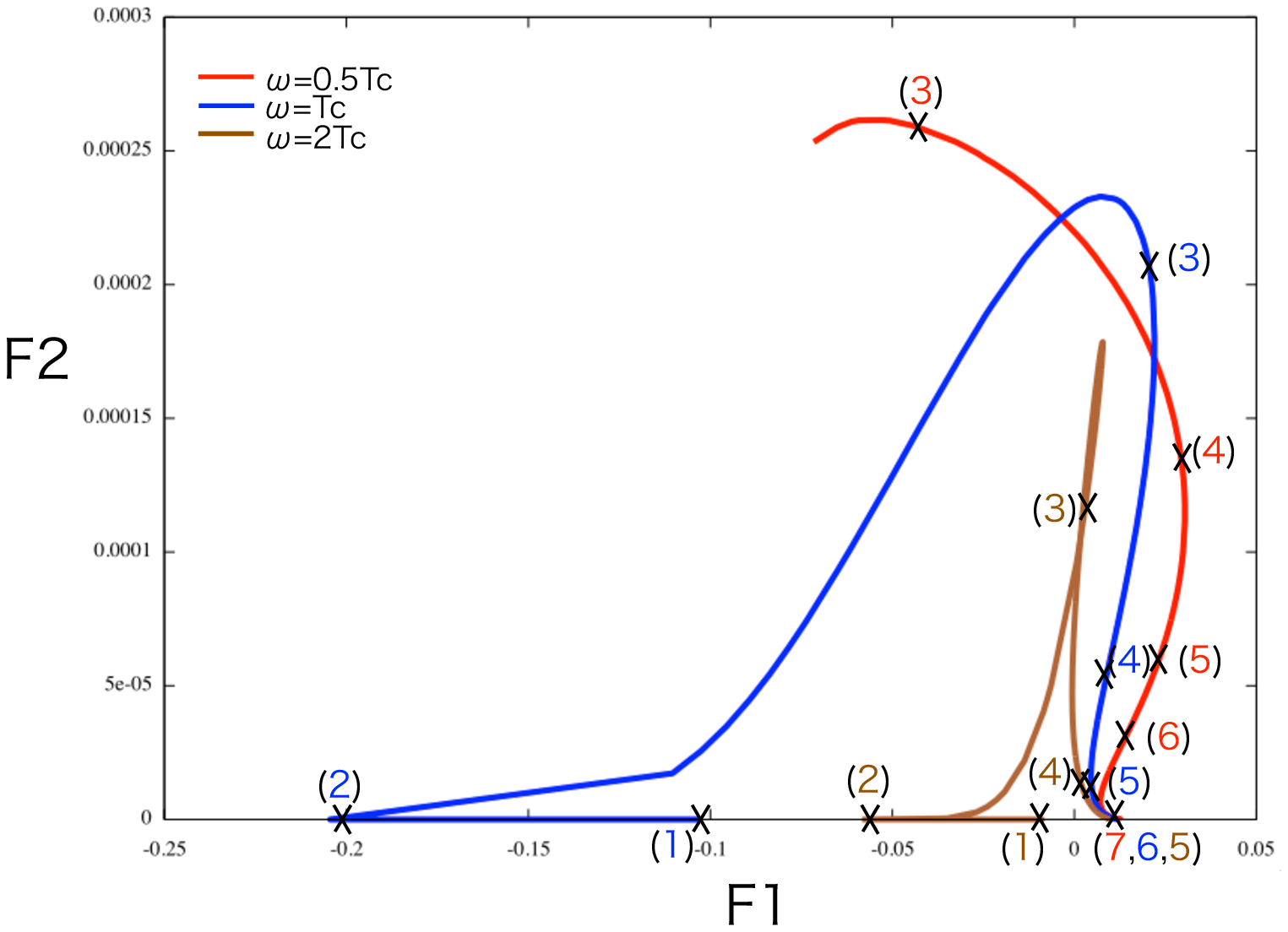}
 \includegraphics[clip,width=110mm]{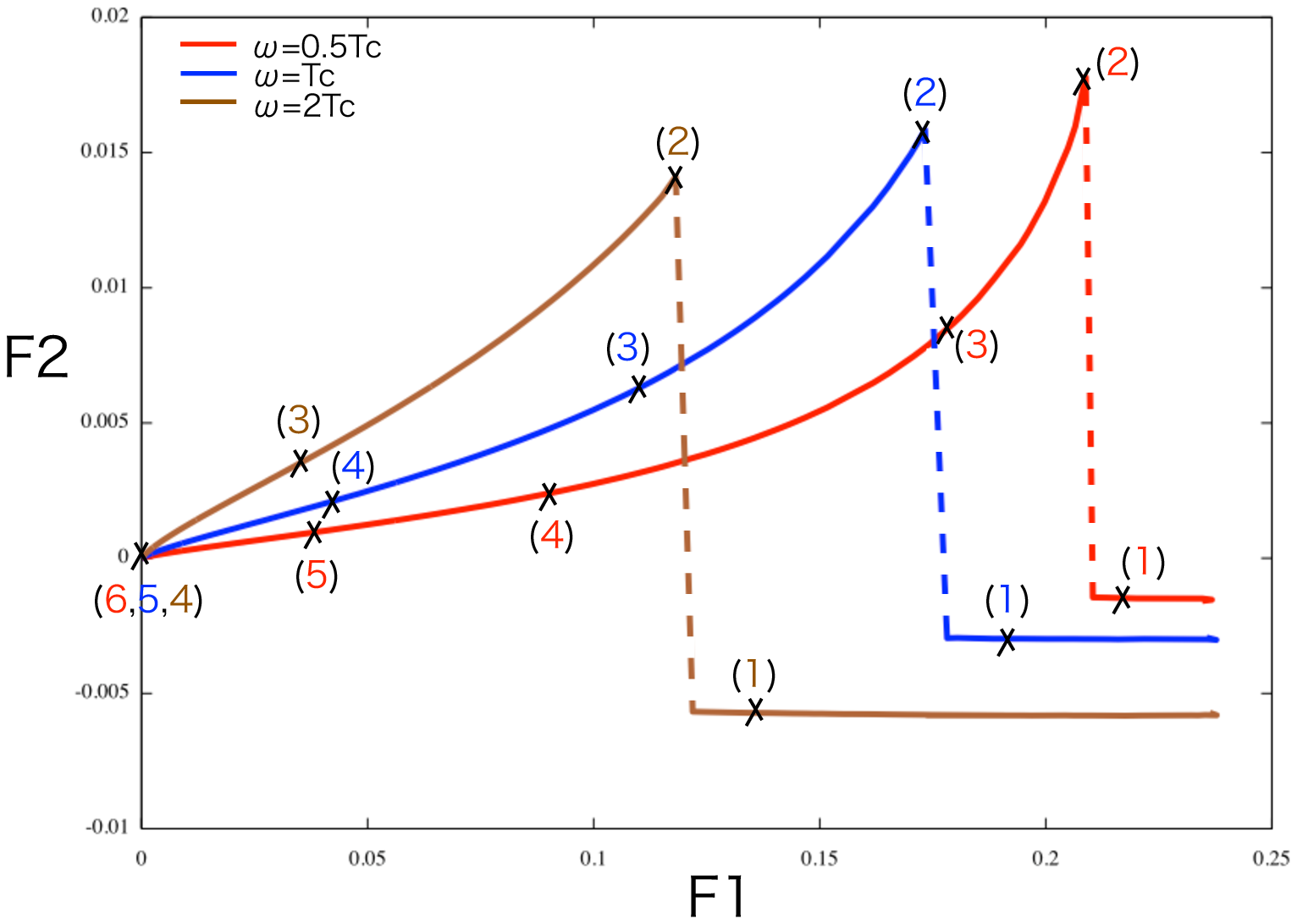}
\end{center}
\caption{Argand plots of $({\cal F}_1 (\omega, q). {\cal F}_2 (\omega, q) )$. The upper panel is calculated at $T= 0.7  T_c$, while the lower panel is at $T=1.4T_c$, each with 3 fixed values of $\omega$ indicated by different colors.   Numbers in the parenthesis correspond to 6 selected values of $q/\omega$ at
(1) 0.8, (2) 1, (3) 2, (4) 5, (5) 10 and (6) 20. }
\label{fig:AG_F}
\end{figure}

In order to see the behavior of $\phi (\omega , q)$, we plot Argand diagrams of $\mathcal{F}(\omega \pm i\epsilon , q)$ in figure \ref{fig:AG_F}. Integral of $\phi (\omega , q)$ determines contribution to pressure of non collective modes. 
As we have seen in subsection \ref{sec:cutoff}, $q$ integral of $\phi $ doesn't have any divergence at large $q$. 
One can see in figure \ref{fig:AG_F} that $\mathcal{F}_2$ becomes small as $q/\omega $ becomes large  
with fixed $\omega $ at both temperatures; actually $\mathcal{F}_2$ and  $\phi (\omega , q)$ become exponentially small at large $q$.

In figure \ref{fig:AG_F},the point marked by (1) is in the time-like region ($\omega >q$) and points marked by (3) or by larger numbers in the parentheses are in the  space-like region ($\omega <q$). 
>From upper panel, we can see that $\phi $ at low temperatures becomes almost $\pi $ in the space-like region because $\phi $ is given by eq.(\ref{eq:phi}). 
It seems that the contribution to pressure from the time-like region is large. However, $\mathcal{F}_2$ in the time-like region contains large vacuum contribution much larger than that of thermal excitations. 
The vacuum pressure will be subtracted so that the contribution from the time-like region becomes small.  Namely,  the effect from finite temperature appears in the space-like region.  
Moreover, the value of $\mathcal{F}_2$ is much smaller than that of $\mathcal{F}_1$. As a result, the value of $\phi $ itself becomes very small. It means that the contribution from non-collective modes is negligible. At high temperature,  the value of $\mathcal{F}_2$  is lager than that at low tempe
 ratures because constituent quark mass becomes zero due to the restoration of the chiral symmetry and because quarks which are suppressed by the Polyakov loop at low temperatures become to excite at high temperatures.  However the value is smaller than that of $\mathcal{F}_1$ so that the value of $\phi $ is small. In addition, a cancelation between positive value of $\phi $ and  negative value of $\phi $ occurs. As a result, the contribution from non-collective modes at hight temperatures is also small. 
%
%\begin{figure}[htbp]
%\begin{center}
%\resizebox*{!}{4.7cm}{
%\includegraphics[clip,width=55mm, angle=270]{q0.7.eps}
%\includegraphics[clip,width=55mm, angle=270]{q1.4.eps}
%\end{center}
%\caption{Left: $\phi $ at T=0.7Tc. Right: $\phi$ at 1.4Tc.}
%\label{fig:F1andF2}
%\end{figure}
% 
% 

\subsubsection{The case with $m_0\neq 0$}

In the case of the chiral limit, we find that the contribution from the collective modes among mesonic excitations are much larger than that of the non-collective modes. 
We anticipate that it remains unchanged  in the case which quarks have finite bare mass, 
violating the chiral symmetry explicitly.
In this subsection, we show that even if quarks have finite mass, collective modes also exist at low temperatures and they disappear at high temperatures. In addition, we show the largest contribution at low temperatures still comes from the collective modes 
and the contribution from the non-collective modes remains negligible 
by calculating integral of argument of complex function 
$\mathcal{M}(\omega \pm i\epsilon , q)$.

%The factor $2$ represents the mass of quark and anti-quark.
When quarks have a finite mass, we cannot separate collective modes from non-collective modes easily
by factorization as in eqs.(\ref{smeson0}) and (\ref{pmeson0}). 
In this case, we first need to determine whether collective modes exist or not. For this reason we calculate the real and imaginary part of $\mathcal{M}(\omega \pm i\epsilon , q)$ separately given by eq.(\ref{phi_of_M}).
%\begin{eqnarray}
%\mathcal{M}_1=\mbox{Re}\bigl[ \frac{1}{2G}-\Pi \bigr] , \ \ 
%\mathcal{M}_2=\mbox{Im}\bigl[ \frac{1}{2G}-\Pi \bigr] 
%\end{eqnarray}
%for pion and for sigma meson respectively.
The conditions of isolated meson poles are given by the vanishing both $\mathcal{M}_1$ and 
$\mathcal{M}_2$ ; in particular the condition $\mathcal{M}_1(\omega , q)=0$ determines the dispersion 
relation of collective meson modes. 

\begin{figure}[htbp]
\begin{center}
%\resizebox*{!}{4.7cm}{
 \includegraphics[clip,width=80mm]{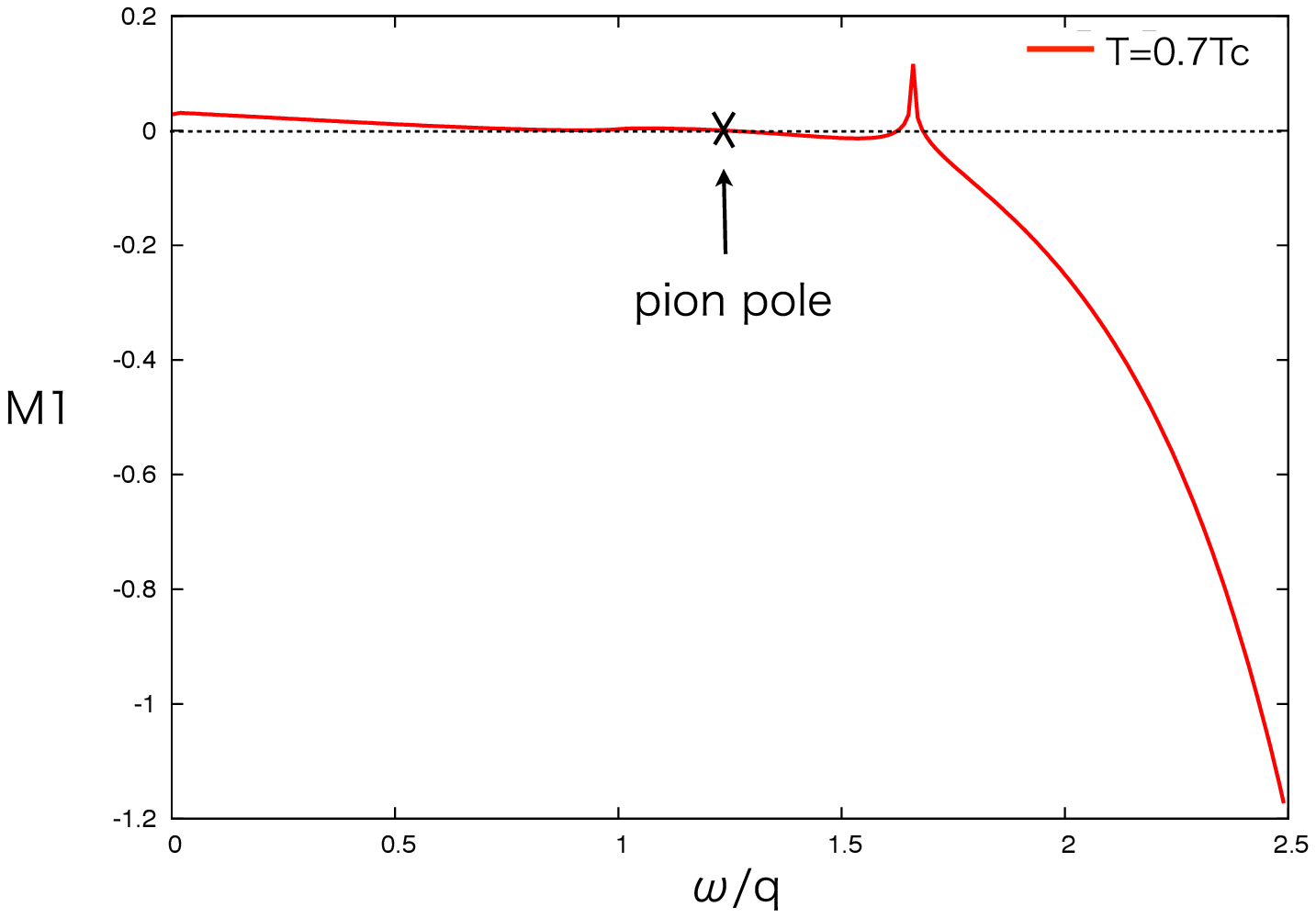}
 \includegraphics[clip,width=80mm]{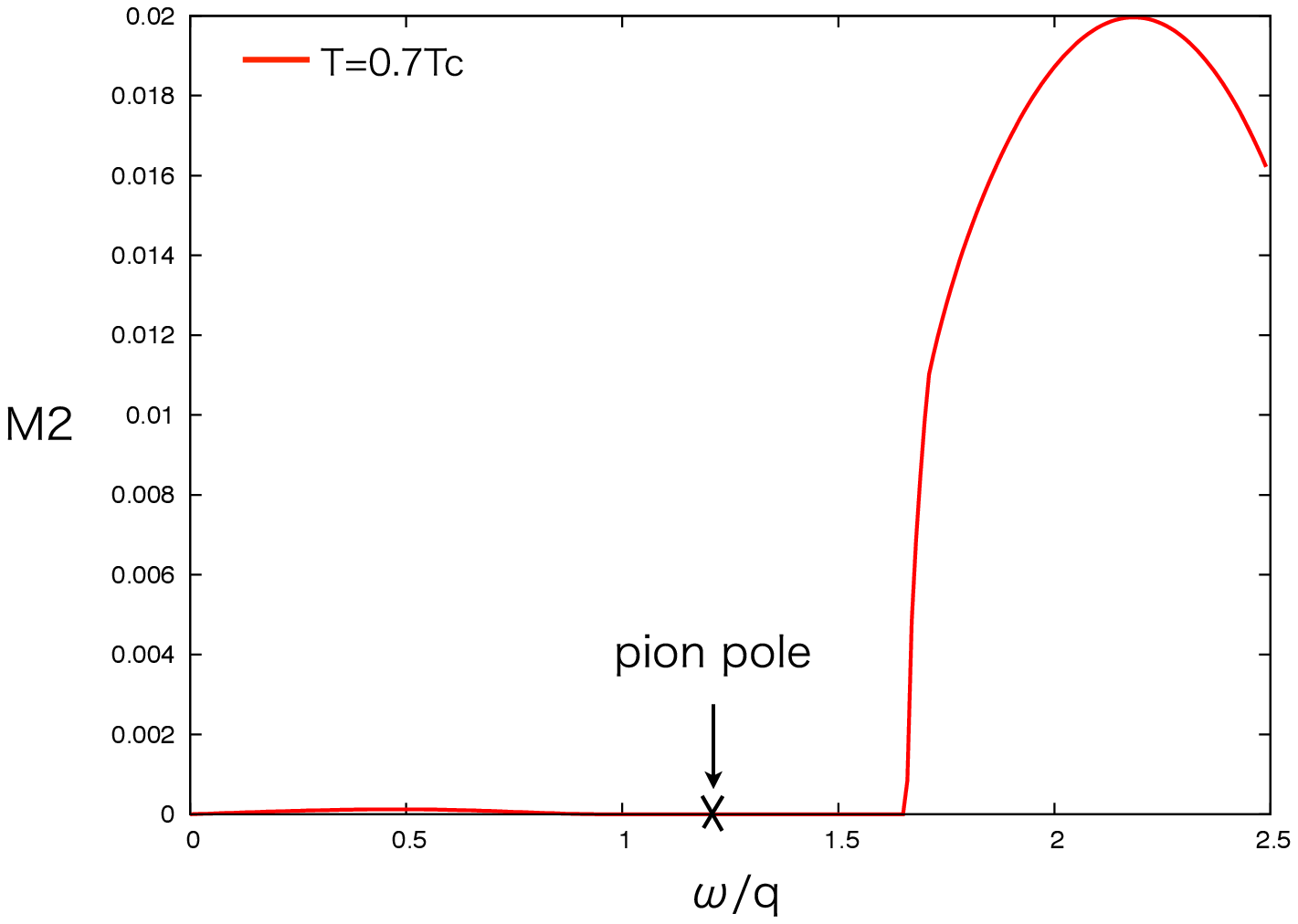}
 \includegraphics[clip,width=80mm]{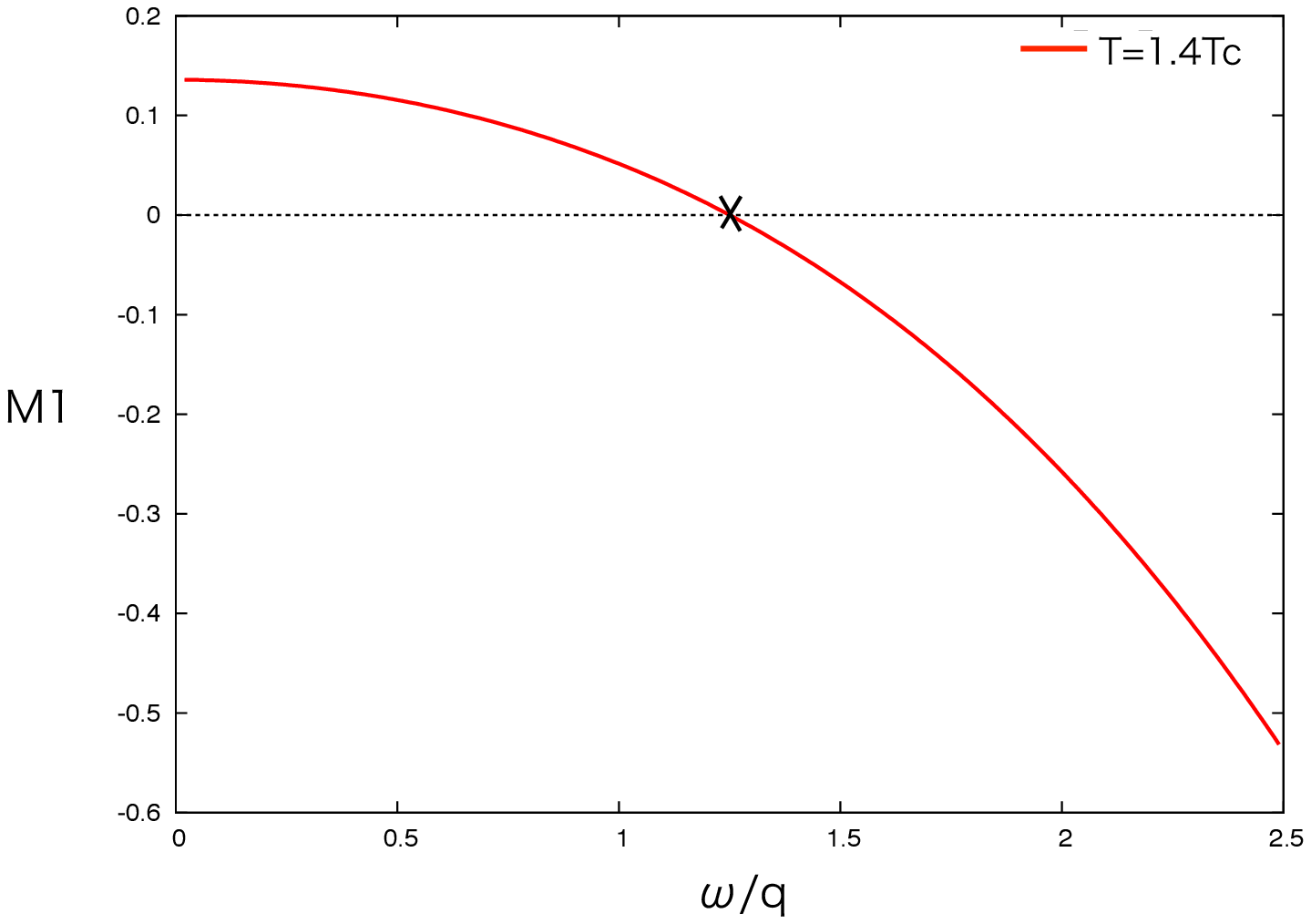}
 \includegraphics[clip,width=80mm]{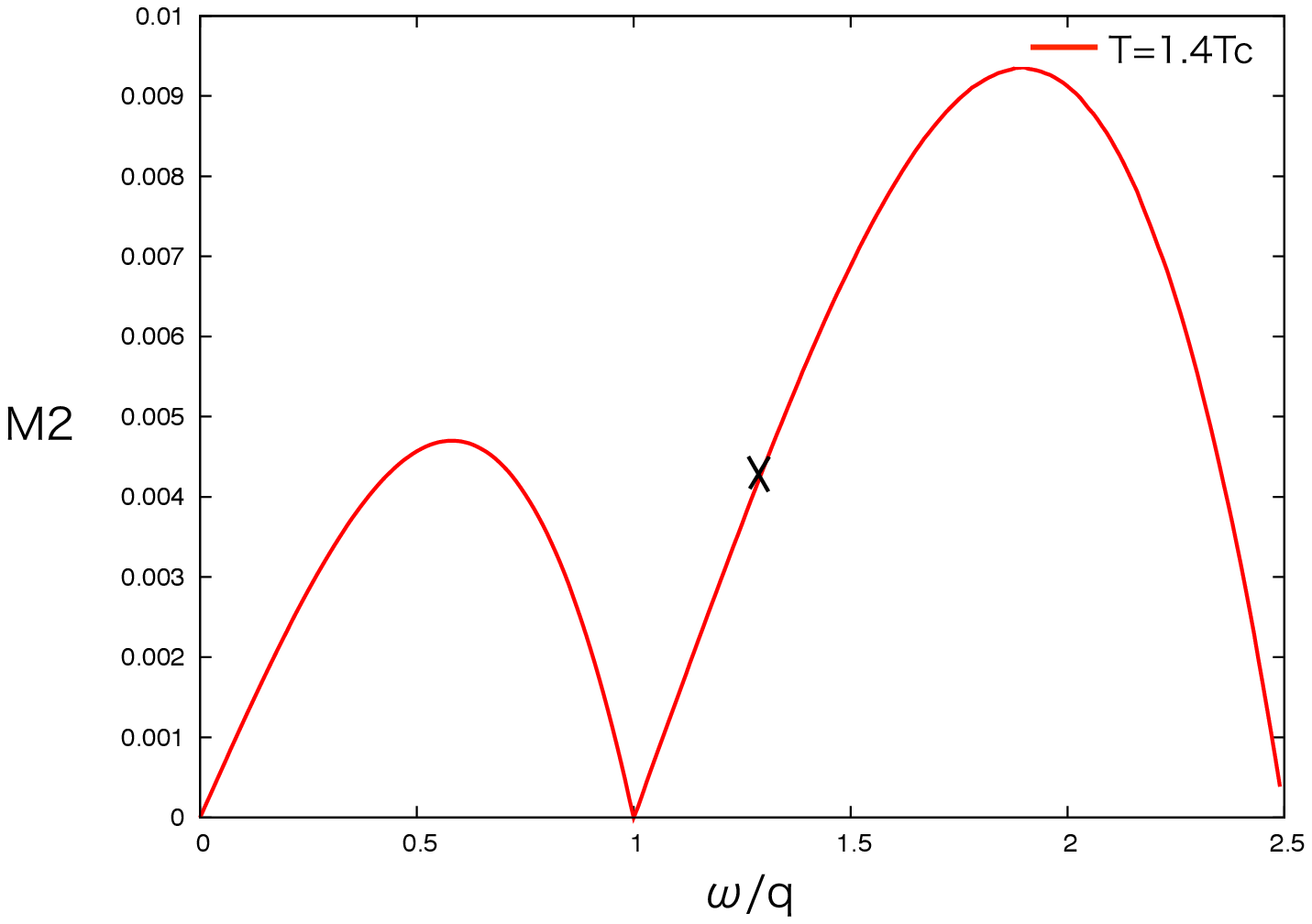}
\end{center}
\caption{M1 and M2 of pion as a function of $\omega $ scaled by q. The upper panels are calculated at 
$T=0.7T_c$ and the lower panels are calculated at $T=1.4T_c$.}
\label{fig:Mpi}
\end{figure}

We show the plots of $\mathcal{M}_1$ and $\mathcal{M}_2$ of pion as a function of $\omega $ scaled by $q$ in figure {\ref{fig:Mpi}}. At low temperatures, $\mathcal{M}_1$ crosses the horizontal axis in the time-like region, and $\mathcal{M}_2$ is also zero at the same value of $\omega /q$ 
where $\mathcal{M}_1$ vanishes. 
It means that a collective pion mode exists at this point.
Furthermore, 
there are no individual excitations in the region where $\mathcal{M}_2 = 0$, since 
 $\mathcal{M}_2$ is proportional to $\mathcal{F}_2 (\omega, q)$, 
%$\mathcal{M}_2 = (-\omega^2 + q^2 ) \mathcal{F}_2 (\omega, q)$ for pions and 
%$\mathcal{M}_2 = (-\omega^2 + q^2 + 4 M_0^2) \mathcal{F}_2 (\omega, q)$ for sigma mesons. 
so that the collective pion modes appear in this region
%not as a resonance with a finite width but 
as stable excitations with infinite lifetime.
On the other hand, at high temperatures, the region where $\mathcal{M}_2=0$ disappears. It means there are no isolated pion poles with infinite lifetime at high temperatures even though $\mathcal{M}_1$ becomes zero.
\begin{figure}[htbp]
\begin{center}
%\resizebox*{!}{4.7cm}{
 \includegraphics[clip,width=80mm]{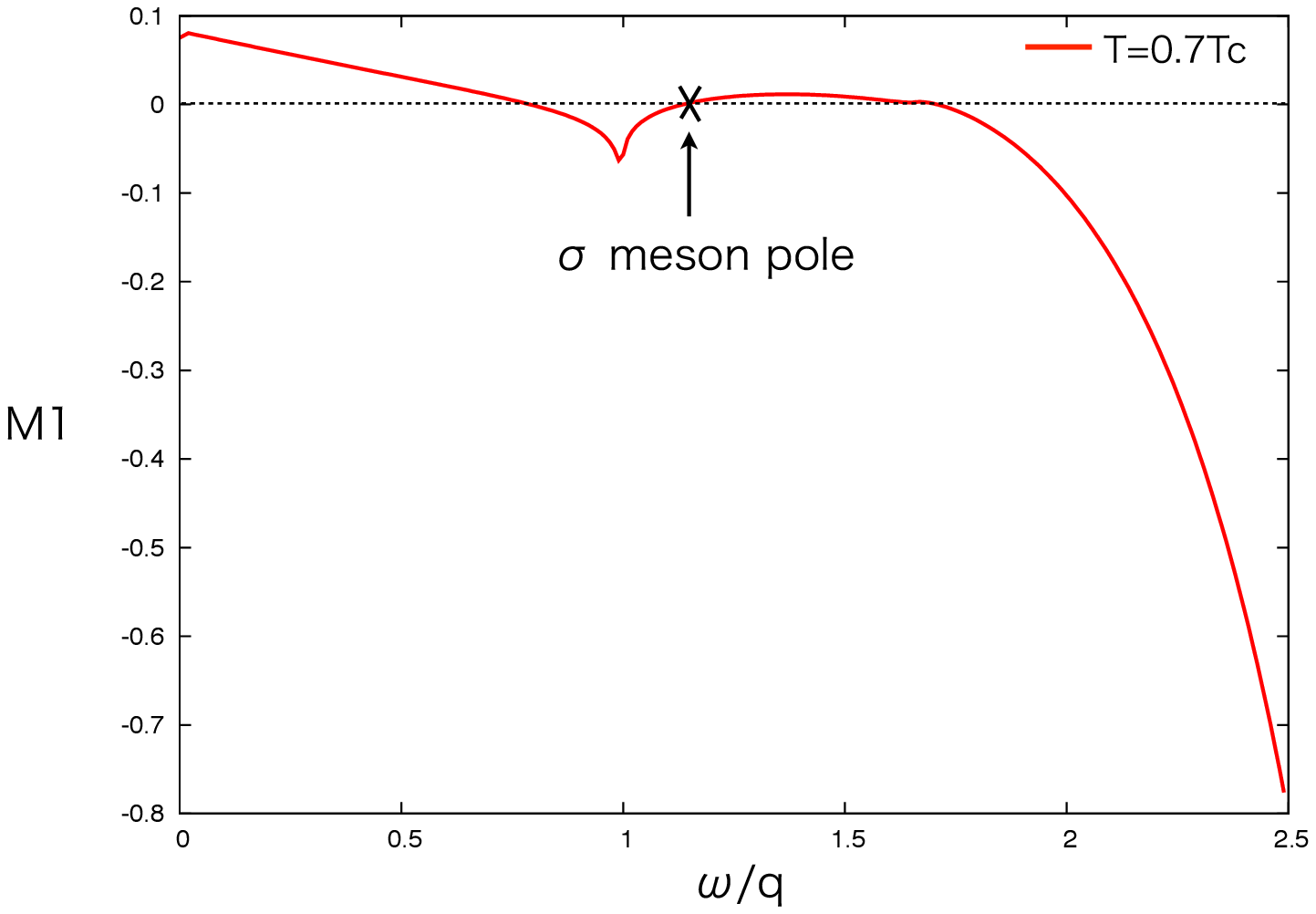}
 \includegraphics[clip,width=80mm]{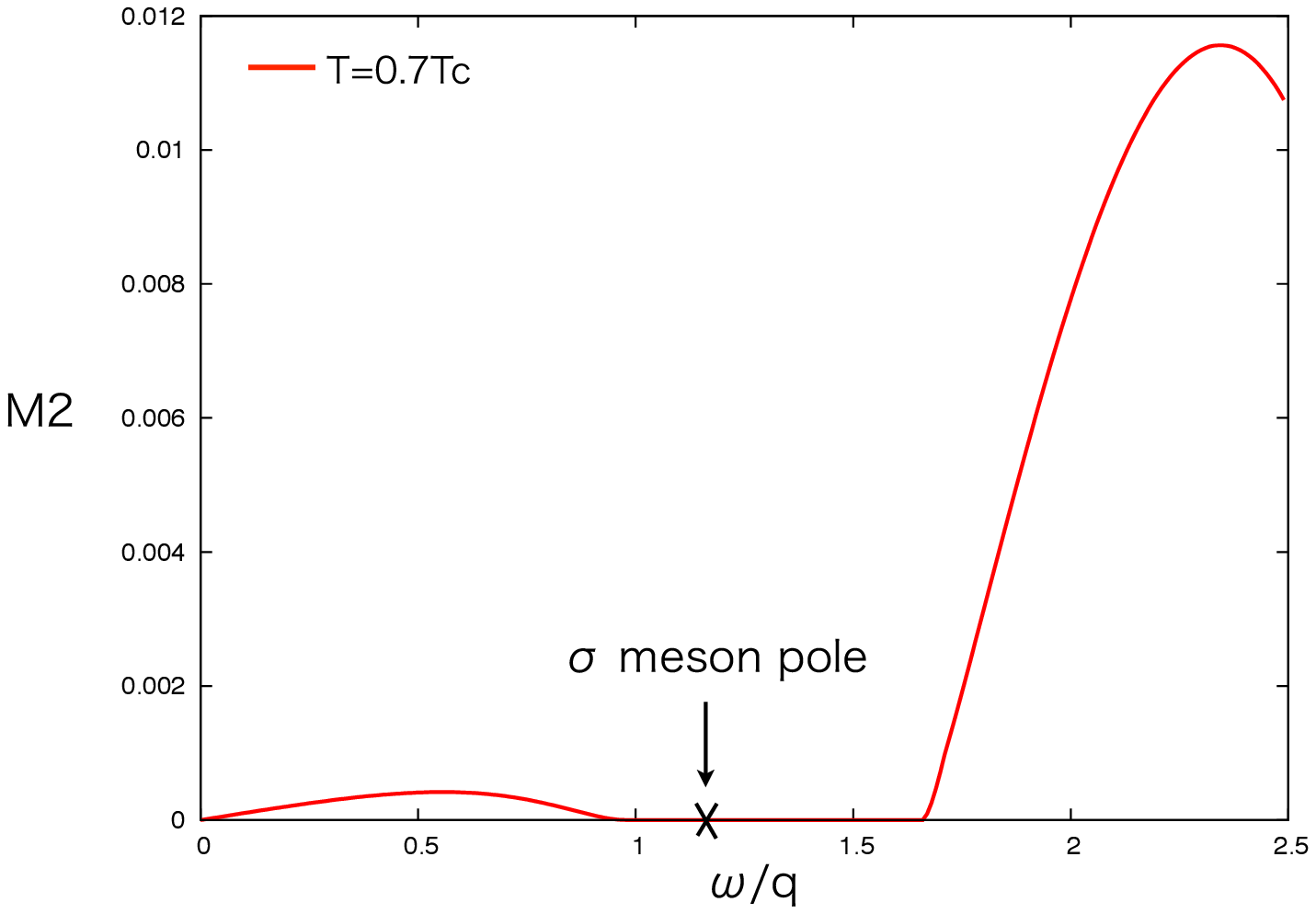}
 \includegraphics[clip,width=80mm]{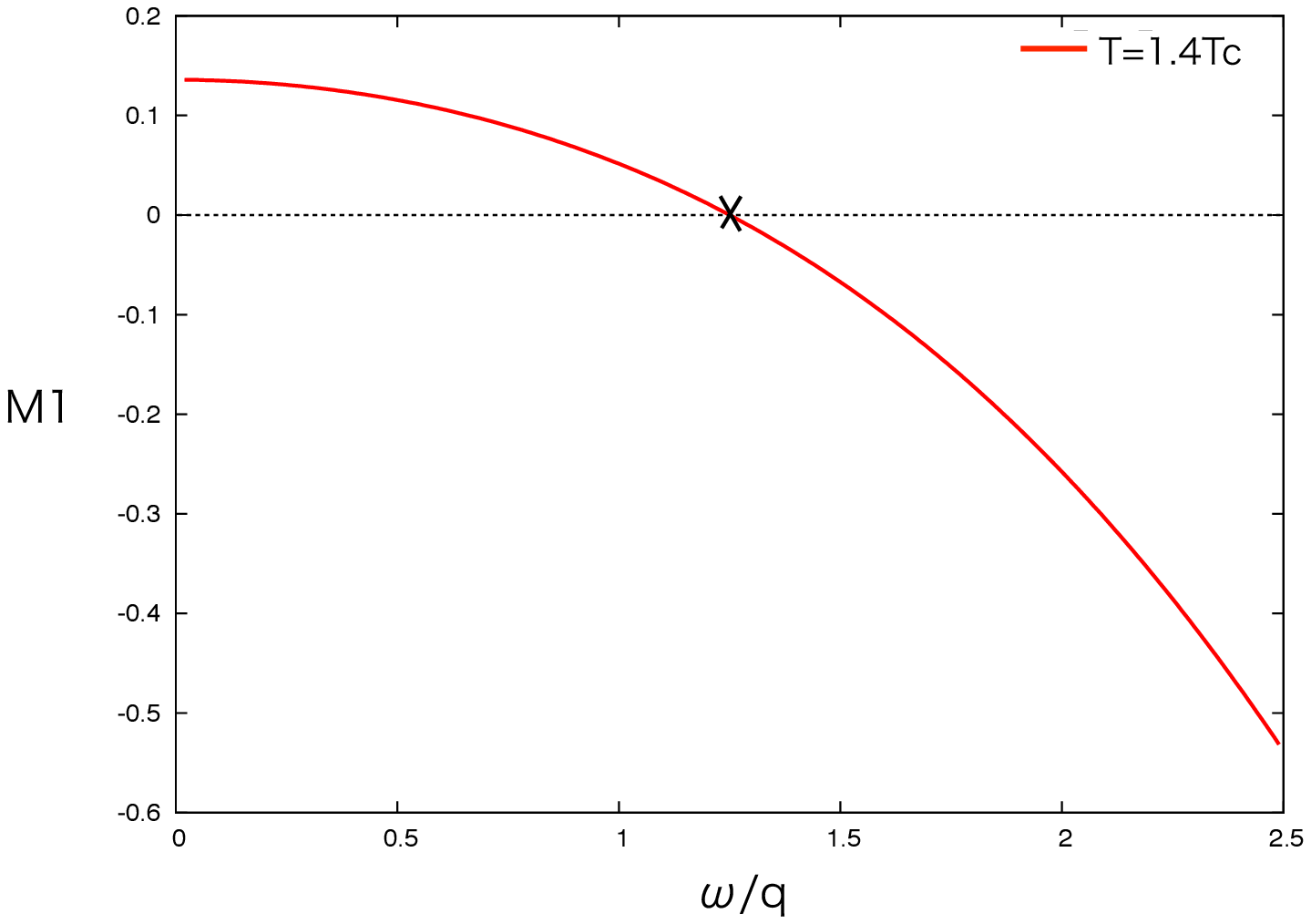}
 \includegraphics[clip,width=80mm]{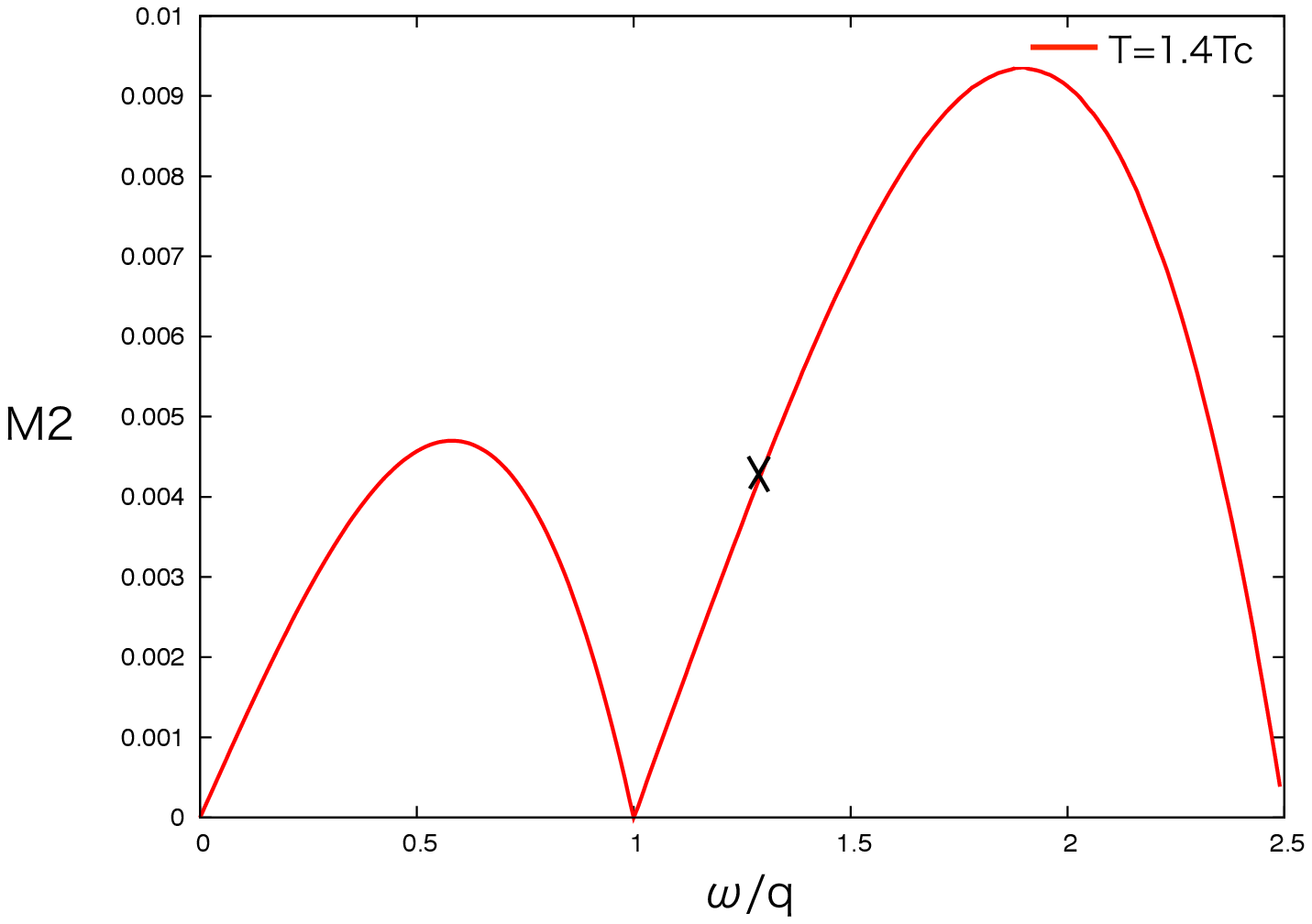}
\end{center}
\caption{M1 and M2 of sigma meson as a function of $\omega $ scaled by q.The upper panels are calculated at 
$T=0.7T_c$ and the lower panels are calculated at $T=1.4T_c$.}
\label{fig:Msi}
\end{figure}
We also show $\mathcal{M}_1$ and $\mathcal{M}_2$ of sigma meson as a function of $\omega $ scaled by $q$ in figure {\ref{fig:Msi}}. The situation is same as the case of pion. At low temperatures there is an  isolated sigma meson pole in the time-like region, while there are no poles at high temperatures.

Next, we consider the contribution from the non-collective modes. 
In the same way as the case in the chiral limit, it is useful to see Argand diagrams of $\mathcal{M}(\omega \pm i\epsilon , q)$ for understanding the contribution to the pressure from non-collective modes.
Note that unlike the case of chiral limit $\mathcal{M}(\omega \pm i\epsilon , q)$ contains not only non-collective modes but also collective modes. However, since collective modes always exist in the region where $\mathcal{M}_2$ is zero, 
we can separate contribution of the collective meson excitations from that of the non-collective modes. 
The pressure from collective meson excitations is given by that of ideal gas with the modified 
dispersion relation determined by $\mathcal{M}_1(\omega,  q)=0$. Therefore the pressure calculated from the integral  of $\phi $ defined by eq.(\ref{phipi_phisigma}) corresponds to the pressure of the non-collective modes.

\begin{figure}[htbp]
\begin{center}
%\resizebox*{!}{4.7cm}{
 \includegraphics[clip,width=110mm]{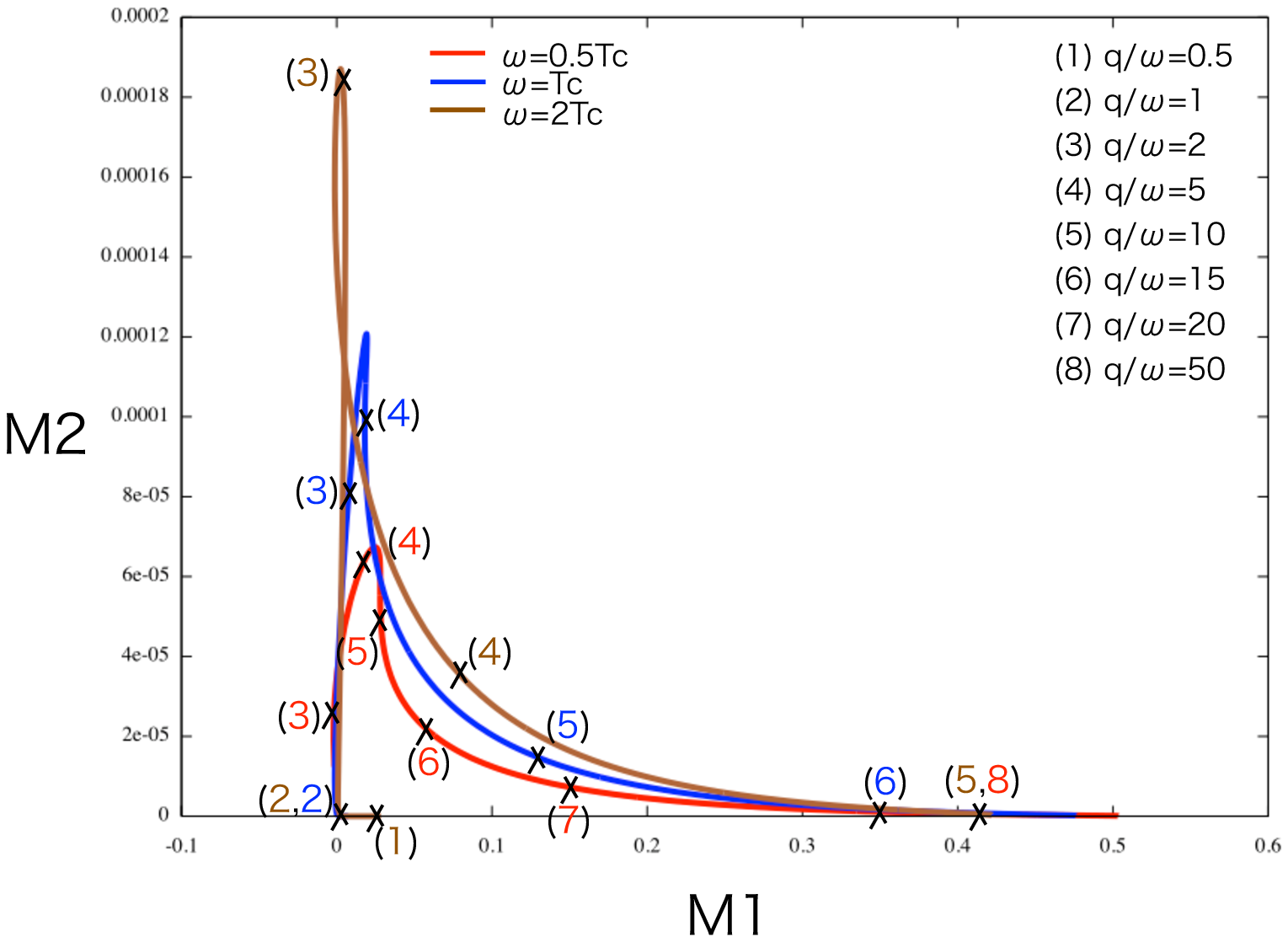}
 \includegraphics[clip,width=110mm]{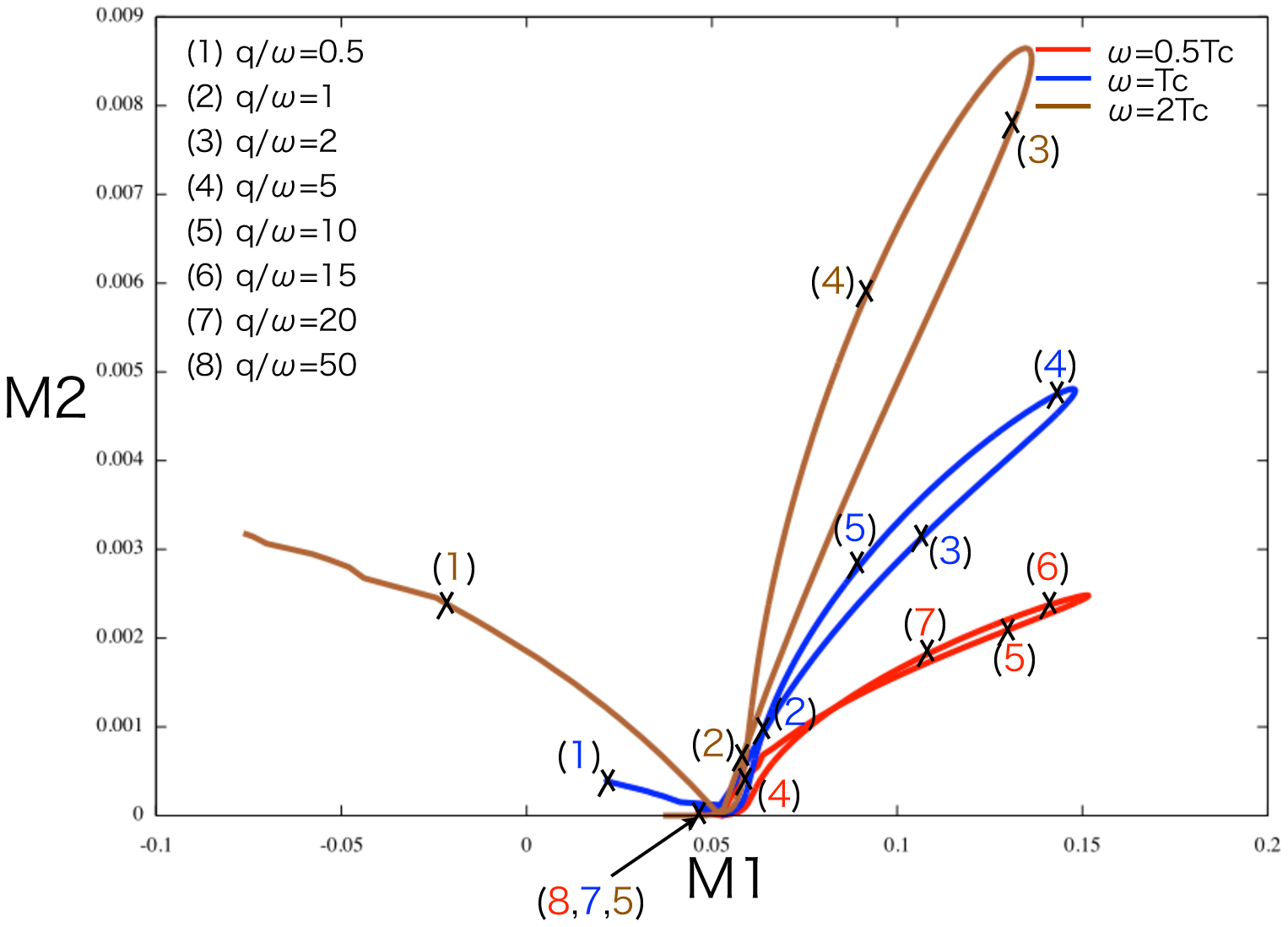}
\end{center}
\caption{Argand diagrams of $\mathcal{M}_{\pi }(\omega +i\epsilon , q)$. The upper panel is computed at $T=0.7T_c$ and the lower panel is cmputed at $T=1.4T_c$.}
\label{fig:AGpi_M}
\end{figure}

\begin{figure}[htbp]
\begin{center}
%\resizebox*{!}{4.7cm}{
 \includegraphics[clip,width=110mm]{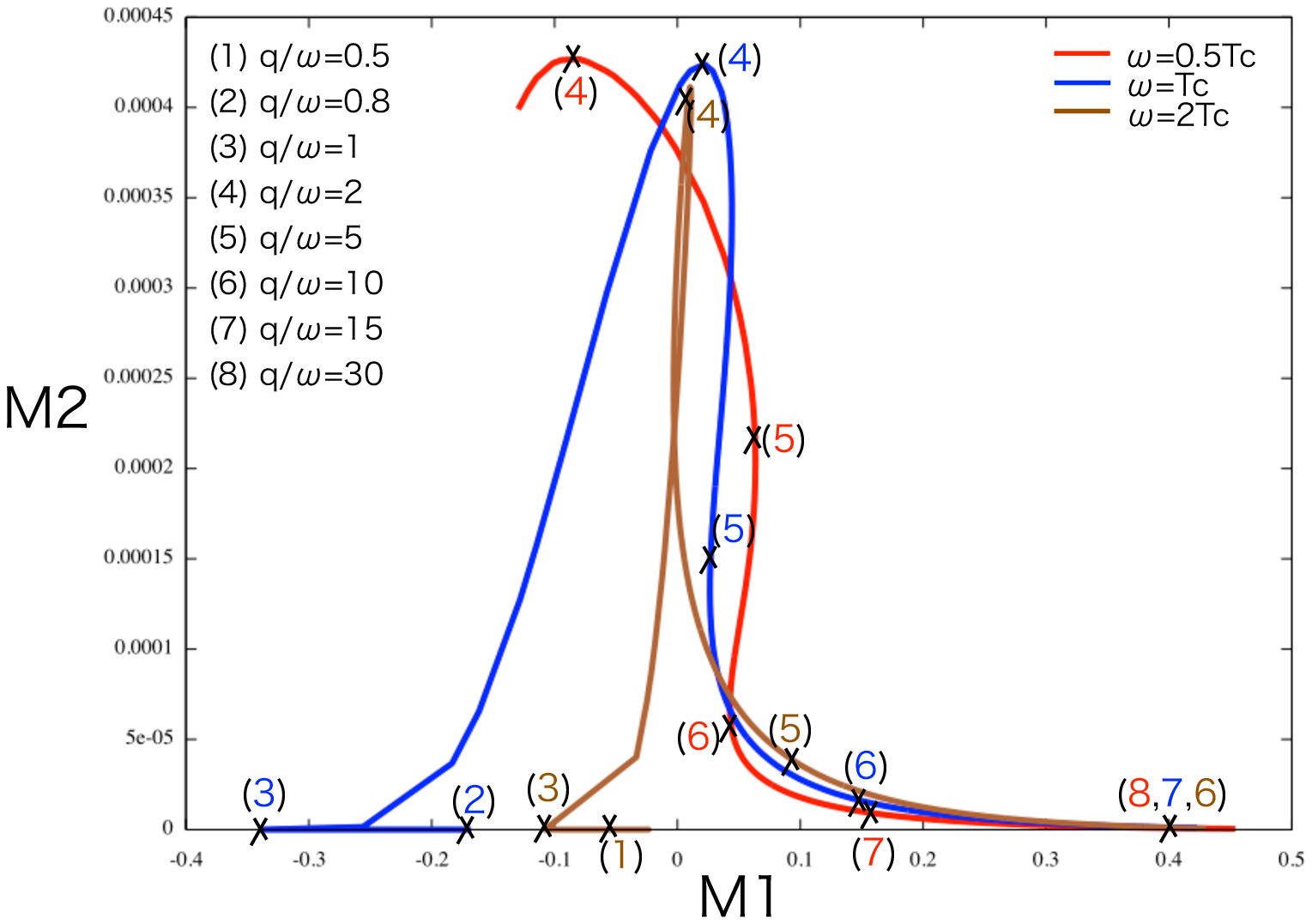}
 \includegraphics[clip,width=110mm]{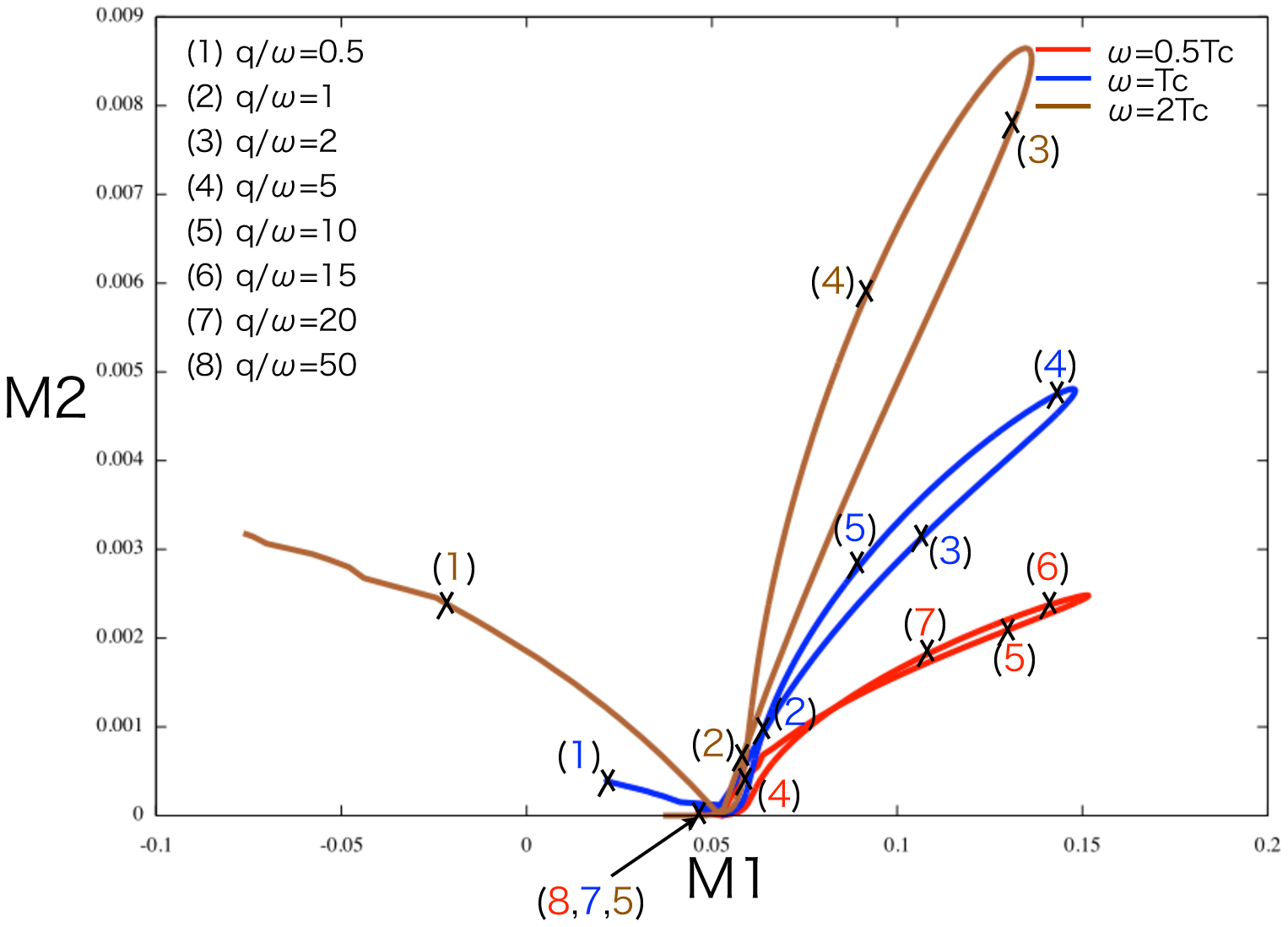}
\end{center}
\caption{Argand diagrams of $\mathcal{M}_{\sigma }(\omega +i\epsilon , q)$.The upper panel is computed at $T=0.7T_c$ and the lower panel is cmputed at $T=1.4T_c$.}
\label{fig:AGsi_M}
\end{figure}

Argand diagrams of $\mathcal{M}_{\pi }(\omega +i\epsilon , q)$ and 
$\mathcal{M}_{\sigma }(\omega +i\epsilon , q)$ are shown in figure \ref{fig:AGpi_M} and figure \ref{fig:AGsi_M} respectively. From these figures, one can see that $\mathcal{M}_2 $ of pion and sigma meson becomes small as $q$ increases.
Actually $\phi $ becomes exponentially  small at large $q$ for a fixed $\omega $ so that the $q$ integral converges. The behavior of $\mathcal{M}_{\pi }(\omega +i\epsilon , q)$ and $\mathcal{M}_{\sigma }(\omega +i\epsilon , q)$ at high temperatures is the same since $M_0=0$ when the chiral symmetry is restored.  
%Since the difference between $\mathcal{M}_{\pi }(\omega +i\epsilon , q)$ and $\mathcal{M}_{\sigma }(\omega +i\epsilon , q)$ is only $M_0$ in factor $(-\omega ^2+q^2)$, there are no difference between them in the region where the chiral symmetry restores.

%
%\begin{figure}[htbp]
%\begin{center}
%\resizebox*{!}{4.7cm}{
 %\includegraphics[clip,width=80mm, angle=270]{phi_omega.eps}
%\includegraphics{PNJLm0.eps}
%\includegraphics{PNJLmf.eps}}
%\end{center}
%\caption{
%The argument of $\mathcal{M}$ as a function of $\omega $ scaled by $q$.
%}
%\label{fig:phase}
%\end{figure}
%
% 
%\begin{figure}[htbp]
%\begin{center}
%\resizebox*{!}{4.7cm}{
 %\includegraphics[clip,width=100mm, angle=270]{phi_BDF.eps}
%\end{center}
%\caption{
%Bosonic distribution function times argument.
%}
%\label{fig:phi}
%\end{figure}
%

\section{Conclusion}

We have studied the quark-hadron phase transition at zero net baryon density 
in terms of the PNJL model which is designed to incorporate both the chiral phase transition and the deconfining transition expected from QCD at finite temperature. 
We have shown that the model describes the low temperature phase as a gas of light mesons 
and the high temperature phase as a gas of free quarks and anti-quarks with restored chiral symmetry.
The transition between two phases is continuous cross-over transition and the thermally excited degrees 
of freedom are transformed from mesonic degrees of freedom to the quark degrees of freedom continuously 
as temperature increases. 

The quark degrees of freedom is suppressed at low temperatures due to the phase cancellation of the color 
gauge fields acting on different color modes of quark excitations, turning the quark distribution functions to 
the distribution of the quark triads whose contribution to the pressure is suppressed exponentially due to the large effective mass.  
The pressure of the low temperature phase is dominated by the mesonic excitations which arise as collective 
mode of excitations of the mean thermal distribution of the quark triads; the contribution of the non-collective 
excitations of quark triads are suppressed.    
As the temperature increases the collective meson modes are absorbed into the continuum of individual excitations.  
They may persist as resonances with finite lifetime due to the Landau-damping.  
We found the contribution of these modes to the pressure is negligible at high temperatures as well.   
In the high temperature deconfining phase, our model gives essentially the same results as the Nambu-Jona-Lasinio model:  the equation of state becomes that of non-interacting massless quark-gluon gas. 

We note that the quark triad is still not a baryon. 
There is no spatial correlation of three quarks since they share exactly the same amount of momentum and energy.    
They also have somewhat over-reduced degrees of freedom.    
It is important to remedy these points for studying further implications of the model 
at finite baryon chemical potentials. 

\section*{Acknowledgements}
We thank G. Baym, J.P. Blaizot, H. Fujii, K. Fukushima, T. Hatsuda, F. Lenz and B. Svetitsky for useful discussions and comments on this work. 
In particular, we thank G. Baym and T. Hatsuda for illuminating discussions on the effective degrees of freedom of quark triads.
KY's work has been supported by the Global COE Program "the Physical Sciences Frontier", MEXT, Japan.   
TM's work has been supported by the Grant-in-Aid \# 21540257
of MEXT, Japan.  

\appendix
\section{Pressure by non-collective mesonic excitations in the chiral limit} 

Here we present the detail of the computation of the pressure exerted by the non-collective mesonic excitation modes given by (5.18). 

We first perform the sum over the bosonic Matsubara frequencies $\omega_n$ in (\ref{noncollective}) by contour integration in the complex z-plane:
\begin{eqnarray}
\Delta p_M(T,A_4) = -2T\int \frac{d ^3 q}{(2\pi)^3} \frac{1}{2\pi i} \int _{\mathcal{C}}dz \frac{1}{e^{\beta z}-1}
\mbox{ln}F( - i z, q, A_4)
\end{eqnarray}
where the contour $\mathcal{C}$ encircles the imaginary axis counter-clock-wise (Fig. A-1). 
The function $1~/~(e^{\beta z}-~1~)$ has poles at $z=2\pi nTi = \omega_n i$ and is analytic everywhere else. 
We change the integration path along the contour $\mathcal{C}$ to the path which encircles real z-axis 
clock-wise (See Fig. A-2).
Denoting the integration variable z by $\omega$ on the real z-axis and then 
converting the integration along the paths sandwiching negative $\omega$ region to the integration
around positive real axis, we find  
\begin{equation}
\Delta p_M (T, A_4) % 4 T \int \frac{d^3 q}{(2\pi)^3} \ln F(\omega_n, q, A_4) 
 = - 2 \int \frac{d ^3 q}{(2\pi)^3} \frac{1}{2\pi i}\int_0^\infty d \omega
\left[ 1 + \frac{2}{ e^{\beta \omega} -1} \right]
\ln \left[ \frac{ {\cal F} ( \omega + i \epsilon, q, A_4)}{ { \cal F} ( \omega- i \epsilon, q, A_4)} \right]
\end{equation}
where we have written
\begin{eqnarray}
{\cal F} ( \omega, q, A_4) & = & F ( - i \omega, q, A_4) \nonumber \\
&  =  & 2 T \sum_{n'} \tr_c \psekibun  
\frac{1}{\left[(\epsilon_{n'} + gA_4)^2 + E_p^2\right] \cdot \left[(\epsilon_{n'} + gA_4 - i \omega)^2 + E_{p+q}^2\right]}  
\end{eqnarray}
where $\epsilon_{n'} = (2n'+1) \pi T$ is the fermionic Matsubara frequencies for quark-quasiparticles. 
If necessary, we may rewrite the above expression by the partial integration, to 
\begin{eqnarray}
\Delta p_M(T,A_4)=-2T\int \frac{d ^3 q}{(2\pi)^3} \frac{1}{2\pi i}\int_0^\infty d \omega
\bigl[ \omega + \frac{2}{\beta }\mbox{ln}(1-\mbox{e}^{\beta \omega })\bigr]\frac{d}{d\omega }
\mbox{ln}\Bigl[ \frac{{ \cal F} (\omega +i\epsilon ,q,A_4)}{{ \cal F} (\omega -i\epsilon ,q,A_4)}\Bigr]  \ \ 
\end{eqnarray}

We now express the sum over the fermionic Matsubara frequencies $\epsilon_{n'}$ by contour integration:
\begin{eqnarray}
{\cal F} ( \omega, q, A_4) & = &   \tr_c \psekibun  \int _{\mathcal{C}} \frac{dz}{2\pi i}  \frac{-1}{e^{\beta z} + 1}
\frac{1}{\left[  ( - (z + i gA_4)^2 + E_p^2\right] \cdot \left[ - (z + i gA_4 + \omega)^2 + E_{p+q}^2\right]}  
\nonumber \\
\end{eqnarray}
where the function $\frac{-1}{e^{\beta z} + 1}$ is chosen so that it gives poles at the fermionic Matsubara frequencies on the imaginary axis; minus sign is necessary for insuring the right sign of the residues at these poles.  The integration path is taken as the same as the bosonic contour integration.  
We change the integration path again to the path enclosing the real z-axis and perform the integral around the poles of the integrand at the real values of z. 
The result may be written as a sum of two terms:
\begin{equation}
{\cal F} (\omega, q, A_4 )  =   {\cal F} _{\rm scat} (\omega, q, A_4 )  +
{\cal F} _{\rm pair} (\omega, q, A_4 ) 
\end{equation}
the first term is scattering term %with space-like energy-momentum $q^2 < \omega^2$,  
 \begin{eqnarray}
{\cal F} _{\rm scat} (\omega, q, A_4  ) & = & \int \frac{d ^3 p}{(2\pi)^3} \frac{1}{2 E_p 2 E_{p+q}} 
\left( \frac{1}{\omega+ E_p - E_{p +q}} - \frac{1}{\omega - E_p + E_{p +q}} \right) 
\nonumber \\
& & \qquad \qquad \times \tr_c \left( f (E_p -igA_4) - f (E_{p+q} - ig A_4 ) \right) 
\end{eqnarray}
while the second term corresponds to the pair creation and annihilation term 
 \begin{eqnarray}
{\cal F} _{\rm pair} (\omega, q,A_4  ) & = & \int \frac{d ^3 p}{(2\pi)^3} \frac{1}{2 E_p 2 E_{p+q}} 
\left( \frac{1}{\omega+ E_p + E_{p +q}} - \frac{1}{\omega - E_p - E_{p +q}} \right) 
\nonumber \\
& & \qquad \qquad \times \tr_c \left( 1 - f (E_p -igA_4) - f (E_{p+q} - ig A_4 ) \right) 
\end{eqnarray}
Again, the external gauge fields appear as a phase factor in the quark distribution function as in the mean-field approximation.  
We therefore replace these phase factors by the Polyakov loops and then substitute them by statistical average 
as in (\ref{averagedis}):
\begin{eqnarray}
{\cal F} ( \omega, q ) & \equiv & \langle{\cal F} ( \omega, q,  A_4 ) \rangle  
= {\cal F}_{\rm scat} (\omega, q) + {\cal F}_{\rm pair} (\omega, q ) 
\end{eqnarray}
where  
 \begin{eqnarray}
 {\cal F}_{\rm scat} ( \omega, q  ) 
%& \equiv & \langle F^{\rm scat} (q, \omega+ i \epsilon, A_4  ) \rangle  \nonumber \\
& = & 3 \int \frac{d ^3 p}{(2\pi)^3} \frac{1}{2 E_p 2 E_{p+q}} 
\left( \frac{1}{\omega+ E_p - E_{p +q}} - \frac{1}{\omega - E_p + E_{p +q}} \right) 
\nonumber \\
& & \qquad \qquad \times  \left( f_\Phi  (E_p) - f_\Phi (E_{p+q}) \right)  \\
{\cal F}_{\rm pair} (\omega, q ) 
\label{Fscat}
%& \equiv  & \langle F^{\rm pair} (q, \omega+ i \epsilon, A_4  ) \rangle \nonumber \\
& = & 3 \int \frac{d ^3 p}{(2\pi)^3} \frac{1}{2 E_p 2 E_{p+q}} 
\left( \frac{1}{\omega + E_p + E_{p +q}} - \frac{1}{\omega - E_p - E_{p +q}} \right) 
\nonumber \\
& & \qquad \qquad \times  \left( 1 - f_\Phi (E_p) - f_\Phi (E_{p+q} ) \right)  
\label{Fpair}
\end{eqnarray}
We may interpret these correlation energy as non-collective fluctuations of the system carrying 
mesonic quantum numbers with quenched quark distributions. 

For the computation of the correlation energy or pressure, it is convenient to decompose the function 
${\cal F} (\omega \pm i \epsilon, q ) $ into real part ${\cal F}_1( \omega, q) $ and imaginary part 
$ {\cal F}_2 ( \omega, q ) $: 
\begin{eqnarray}
{\cal F} ( \omega \pm i \epsilon, q ) & = & {\cal F}_1( \omega, q) \pm i {\cal F}_2 (\omega, q ) 
%\nonumber \\
%{\cal F} ( \omega \pm i \epsilon, q ) =| {\cal F} (q, \omega \pm i \epsilon ) | e^{\pm i \phi (q, \omega) } 
=  \sqrt{  {\cal F}_1( \omega, q )^2 +  {\cal F}_2 ( \omega, q )^2 } e^{\pm i \phi ( \omega, q) }  ,
\end{eqnarray}
%with the modulus and the argument given by
%\begin{eqnarray}
%& & | {\cal F} (q, \omega \pm i \epsilon ) | = \sqrt{  {\cal F}^1(q, \omega)^2 +  {\cal F}^2 (q, \omega )^2 } \\
%\end{eqnarray}
where the argument $\phi$ is given by
\begin{equation}
 \phi (\omega, q) =  \tan^{-1} \frac{{\cal F}_2(\omega, q)}{{\cal F}_1(\omega, q)} ~~. \label{argumentofF}
\end{equation}
The real part and imaginary part of the function ${\cal F}$ are further decomposed into two parts: 
the scattering term and the pair excitation term.
The two components of the real part are given by the principal part integrals: 
\begin{eqnarray}
{\cal F}_{\rm scatt., 1} (\omega, q) %& = & Re F(q, \omega + i \epsilon) \nonumber \\
& = & {\cal P} \int \frac{d ^3 p}{(2\pi)^3} \frac{1}{2 E_p 2 E_{p+q}}
\left( \frac{1}{\omega + E_p - E_{p +q}} - \frac{1}{\omega - E_p + E_{p +q}} \right) 
\\
& & \hbox{\hskip 5cm}  \times 
\left( f_\Phi (E_p) - f_\Phi (E_{p+q} ) \right) \nonumber \\ 
{\cal F}_{\rm pair, 1} ( \omega, q) %& = & Re F(q, \omega + i \epsilon) \nonumber \\
& = & {\cal P} \int \frac{d ^3 p}{(2\pi)^3} \frac{1}{2 E_p 2 E_{p+q}}
%& &  \left. \qquad \qquad \qquad \qquad +  
\left( \frac{1}{\omega + E_p + E_{p +q}} - \frac{1}{\omega - E_p - E_{p +q}} \right)
\nonumber \\
& & \hbox{\hskip 5cm}  \times  
\left( 1 - f_\Phi (E_p) - f_\Phi (E_{p+q} ) \right) 
% \right] \nonumber \\
\end{eqnarray}
while the two components of the imaginary part contains the energy conserving $\delta$-functions:
\begin{eqnarray}
{\cal F}_{\rm scatt., 2} (\omega, q) %& = &  Im F(q, \omega + i \epsilon) \nonumber \\ 
& = &  - \pi  \int \frac{d ^3 p}{(2\pi)^3} \frac{1}{2 E_p 2 E_{p+q}}
\left( f_\Phi (E_p) - f_\Phi (E_{p+q} ) \right)
\nonumber \\
& & \qquad \qquad \qquad \times \left( \delta (\omega + E_p - E_{p +q}) - \delta ( \omega - E_p + E_{p +q} ) \right) 
 \nonumber \\ 
{\cal F}_{\rm pair, 2} (\omega, q) 
& = &  - \pi  \int \frac{d ^3 p}{(2\pi)^3} \frac{1}{2 E_p 2 E_{p+q}} %\left. \qquad \qquad \qquad \qquad +  
\left( 1 - f_\Phi (E_p ) - f_\Phi (E_{p+q} ) \right) 
%\left( \frac{1}{\omega + E_p + E_{p +q}} - \frac{1}{\omega - E_p - E_{p +q}} \right)
\nonumber \\
& & \qquad \qquad \qquad \times \left( \delta (\omega + E_p + E_{p +q}) - \delta ( \omega - E_p - E_{p +q} ) \right) 
% \right] \nonumber \\
\end{eqnarray}
Similar function for the finite temperature bosonic system has been computed\cite{Matsui:2008ei, Matsuo:2008cm}.
It is evident that the scattering term has non-zero imaginary term in the space-like energy-momentum region 
($\bq^2 > \omega^2$), while the pair creation/annihilation term is non-vanishing only in the time-like region
($\bq^2 < \omega^2$).  
It is important to note that non-collective mesonic correlation arises only from non-vanishing 
imaginary part of ${\cal F} (q, \omega)$.

%%%%%%%%%%%%%%%%%%%%%%%%%%%%%%%%%%%

\section{Computation of ${\cal F}_{\rm scat} (\omega, q) $ and ${\cal F}_{\rm pair} (\omega, q) $}
Here we present the computation of the phase space integral 
%To perform the phase space integrations 
in ${\cal F}_{\rm scat} (\omega, q) $ and ${\cal F}_{\rm pair} (\omega, q) $.

\begin{figure}[htbp]
\begin{center}
  \includegraphics[clip,width=80mm]{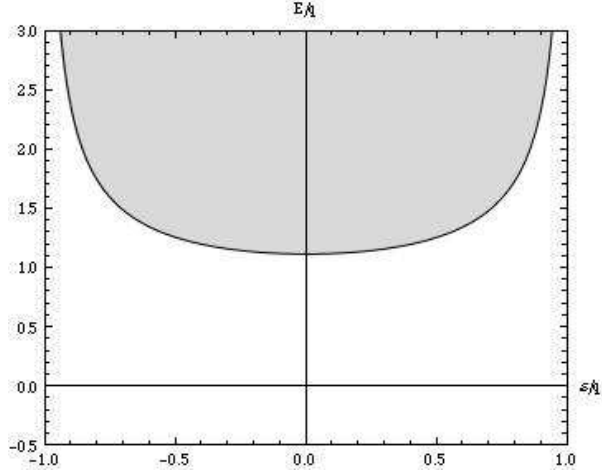}
\end{center}
\caption{Region of integration on $(\varepsilon, E)$ plane is shown by gray area.}
\label{ApB}
\end{figure}

We first adopt the polar coordinate system, $d^3 p= dp p^2 d \cos \theta d \phi$, and then change the variables $(p, \cos \theta) $ to new variables $(E, \varepsilon)$ defined by
\begin{eqnarray}
E & = & \frac{1}{2} ( E_p + E_{p + q} ) \\ 
 \varepsilon & = & E_{p + q} - E_p .
\end{eqnarray}
Note that $2E$ represents the energy absorbed by pair excitation while $\varepsilon$ is the energy transfered by scattering. 
The Jacobian of this transformation is 
\begin{equation}
\frac{\partial (E, \varepsilon)}{\partial (p, \cos \theta) } = \frac{p^2q}{E_p E_{p+q} }
\end{equation}
so that the integration measure absorbs the Lorentz factor $E_p E_{p+q}$:
\begin{equation}
 \frac{d ^3 p}{E_p E_{p+q}} 
 = \frac{dp p^2 d \cos \theta d \phi}{ E_p E_{p+q}} 
 = \frac{1}{ q} dE d \varepsilon d \phi
\end{equation}
Note that by construction
\begin{equation}
2 E \varepsilon = E_{p+q}^2 - E_p^2 = 2 p q \cos \theta + q^2 
%= 2 \sqrt{(E - \frac{1}{2}\varepsilon)^2 - m^2} q \cos \theta + q^2 
\end{equation}
so that the domain of the integration on the $(E, \varepsilon)$ plane is constrained by the condition
\begin{equation}
\cos^2 \theta = \frac{E \varepsilon - \frac{q^2}{2}}{pq} \le 1
\end{equation}
Since $p = \sqrt{E_p^2 - m^2}= \sqrt{(E - \frac{1}{2}\varepsilon)^2 - m^2} $, this implies 
\begin{equation}
E^2 \varepsilon^2 \le q^2 \left( E^2 +\frac{1}{4}\varepsilon^2 - m^2 - \frac{q^2}{4} \right )
\label{boundary}
\end{equation}
The region of the integration on $(E, \varepsilon)$ plane is indicated by gray zone in Fig. \ref{ApB}.

We can now rewrite (\ref{Fscat}) and (\ref{Fpair})  in simpler, physically more transparent  forms.  
The scattering term (\ref{Fscat}) may be written as
\begin{eqnarray}
 {\cal F}_{\rm scat} ( \omega, q  ) 
%& \equiv & \langle F^{\rm scat} (q, \omega+ i \epsilon, A_4  ) \rangle  \nonumber \\
%& = &  \frac{3}{4}  \frac{1}{(2\pi)^2 q}  \int_{-q}^q d \varepsilon  
%\left( \frac{1}{\omega - \varepsilon} - \frac{1}{\omega + \varepsilon } \right) \int_{E_0 (\varepsilon)} d E
%\left( f_\Phi  (E - \frac{1}{2} \varepsilon) - f_\Phi (E +  \frac{1}{2}\varepsilon) \right) 
%\nonumber \\
& = & \frac{1}{2\pi} \int_{-q}^q d \varepsilon  
\left( \frac{1}{\omega - \varepsilon} - \frac{1}{\omega + \varepsilon } \right) N_{\bf scat} (\varepsilon) 
\label{Fscat0}
\end{eqnarray}
where
\begin{eqnarray}
 N_{\bf scat} (\varepsilon, q) & = & \frac{1}{2\pi q} \int_{E_0 } d E
 \left( f_\Phi  (E - \frac{1}{2} \varepsilon) - f_\Phi (E +  \frac{1}{2}\varepsilon) \right) 
 \label{Nscat}
\end{eqnarray}
may be interpreted as the number of scattering states with energy-momentum transfer $(\varepsilon, q)$. 
The lower boundary of the $E$-integration, $E_0$, is given from the constraint (\ref{boundary}) 
as a function of $(\varepsilon, q)$:  
\begin{equation}
E_0 ({\varepsilon}, q) = \frac{q}{2} \sqrt{1 + \frac{4m^2}{q^2 - \varepsilon^2}}
\end{equation}
If the condition $\varepsilon << q$ is fulfilled,  
the integration may be performed approximately as 
\begin{eqnarray}
 N_{\bf scat} (\varepsilon, q) 
\simeq - \frac{1}{2\pi} \frac{\omega}{q} \int_{E_0} dE \frac{d f_\Phi (E)}{dE}
= \frac{1}{2\pi} \frac{\omega}{q} f_\Phi ( E_0 ) \label{N_scat}
\end{eqnarray}
Note that the number of scattering final states decreases exponentially for large $q$. 
Similarly we can express the pair excitation term (\ref{Fpair}) as
\begin{eqnarray}
 {\cal F}_{\rm pair} ( \omega, q  ) 
& = &  \frac{1}{2\pi} \int_{E_0} d E 
\left( \frac{1}{\omega - 2E} - \frac{1}{\omega + 2E } \right) N_{\bf pair} (E, q) 
\label{Fpair0}
\end{eqnarray}
where
\begin{equation}
 N_{\bf pair} (E, q)  =  \frac{1}{2\pi q} \int_{\varepsilon_- }^{\varepsilon_+} d \varepsilon
 \left( 1 - f_\Phi  (E - \frac{1}{2} \varepsilon) - f_\Phi (E +  \frac{1}{2}\varepsilon) \right) 
\end{equation}
may be interpreted as the number of pair excitation states with energy-momentum  $(2E, q)$.  
The boundary of the $\varepsilon$-integration is obtained from (\ref{boundary}) as a function of $E$ and $q$:
\begin{equation}
\varepsilon_\pm (E, q ) = \pm q \sqrt{1- \frac{m^2}{E^2- q^2/4}}
\end{equation}
Since the value of $\varepsilon$ is constrained by $- q < \varepsilon < q$, we can estimate the integration approximately 
and find
\begin{equation}
 N_{\bf pair} (E, q)  \simeq  \frac{1}{2\pi q} ( \varepsilon_+ - \varepsilon_- ) \left(1 + 2 f_\Phi  (E) \right) 
 = \frac{1}{\pi} \sqrt{1- \frac{m^2}{E^2- q^2/4}} \left( 1 - 2 f_\Phi  (E) \right) 
\end{equation}
 
The functions $N_{\bf scat} (\varepsilon) $ and $N_{\bf pair} (E)$ determine the imaginary part of 
${\cal F}_{\rm scat} (\omega+ i \epsilon , q) $ and ${\cal F}_{\rm pair} (\omega+ i \epsilon, q) $, respectively.
Using the expression (\ref{Fscat0}), we find
\begin{equation}
{\cal F}_{\rm scat, 2} (\omega, q) = \frac{1}{2\pi} \int_{-q}^q d \varepsilon  
\left(  - \pi \delta (\omega - \varepsilon) + \pi \delta (\omega + \varepsilon) \right) N_{\bf scat} (\varepsilon) 
 = - N_{\bf scat} (\omega) \label{F_scat2}
\end{equation}
where we have used that $N_{\bf scat} (\varepsilon)$ defined by (\ref{Nscat}) is an odd function of $\varepsilon$ when extended to the negative value of $\varepsilon$. 
Similarly, from (\ref{Fpair0}) we obtain
\begin{equation}
{\cal F}_{\rm pair, 2} (\omega, q) = \frac{1}{2\pi} \int_{E_0} d E
\left(  - \pi \delta (\omega - 2E) + \pi \delta (\omega + 2E) \right) N_{\bf pair} (E ) 
 = - N_{\bf pair} (\omega / 2) 
\end{equation}

%%%%%%%%%%%%%%%%%%%%%%%%%%%%%%%%%%%%%%%%

\section{Separation of collective mesonic modes with finite bare quark mass}
In the presence of non-vanishing bare quark mass, the collective meson modes are also contained in (\ref{Pmm}) or
\begin{eqnarray}
\Delta p_M (T, A_4) % 4 T \int \frac{d^3 q}{(2\pi)^3} \ln F(\omega_n, q, A_4) 
 & = & -  \int \frac{d ^3 q}{(2\pi)^3} \frac{1}{2\pi i}\int_0^\infty d \omega
\left[ 1 + \frac{2}{ e^{\beta \omega} -1} \right]  \nonumber \\ 
& & \qquad \qquad \times \left\{ 3\ln \left[ \frac{{\cal M}_\pi( \omega + i \epsilon, q)}{{\cal M}_\pi ( \omega- i \epsilon, q)} \right] 
+\ln \left[ \frac{{\cal M}_\sigma( \omega + i \epsilon, q)}{{\cal M}_\sigma( \omega- i \epsilon, q)} \right] 
 \right\}
\end{eqnarray}
together with non-collective excitation modes. 
In this case we can still separate the contribution from the collective meson modes.

We first note that for small, but finite $\epsilon$, 
\begin{eqnarray}
 {\cal M}_{\pi}^1 (\omega, q) 
 & = &  ( - \omega^2 + \bq^2) {\cal F}_1 (\omega, q) + {\displaystyle \frac{m_0}{2G M_0} } 
  \label{Mpi}\\
 {\cal M}_{\sigma}^1 (\omega, q ) 
 & = &  ( - \omega^2 + \bq^2 + 4 M_0^2) {\cal F}_1 (\omega, q) + {\displaystyle \frac{m_0}{2G M_0} } 
 \label{Msigma}
\end{eqnarray}
while 
\begin{eqnarray}
 {\cal M}_{\pi}^2 (\omega, q) 
 & = &  - 2 \omega \epsilon {\cal F}_1 (\omega, q) + ( - \omega^2 + \bq^2) {\cal F}_2 (\omega, q) 
 \label{Mpi}\\
 {\cal M}_{\sigma}^2 (\omega, q ) 
 & = &   - 2 \omega \epsilon  {\cal F}_1 (\omega, q) + ( - \omega^2 + \bq^2 + 4 M_0^2) {\cal F}_2 (\omega, q) 
 \label{Msigma}
 \end{eqnarray}
so that in the kinematical region where no individual excitation exists, ${\cal F}_2 (\omega, q) = 0$, we have
\begin{eqnarray}
 {\cal M}_{\pi/\sigma}^2 (\omega, q ) 
 & = &   - 2 \omega \epsilon  {\cal F}_1 (\omega, q) 
 \end{eqnarray}
Since ${\cal M}_{\pi/\sigma}^1 = 0$ at meson poles, the argument of 
${\cal M}_{\sigma} (\omega + i \epsilon, q )$ becomes $\pm \pi/2$, depending only on the sign of ${\cal F}_1(\omega, q)$.
In the limit $\epsilon \to 0$, the poles move to the real $\omega$ axis.  

The locations of the meson poles on $(\omega, q)$ plane are determined by the conditions:
\begin{eqnarray}
 {\cal M}_{\sigma}^1 (\omega, q ) & = &  ( - \omega^2 + \bq^2 + 4 M_0^2) {\cal F}_1 (\omega, q) + {\displaystyle \frac{m_0}{2G M_0} } = 0 
 \label{Msigma}\\
 {\cal M}_{\pi}^1 (\omega, q) & = &  ( - \omega^2 + \bq^2) {\cal F}_1 (\omega, q) + {\displaystyle \frac{m_0}{2G M_0} } = 0
 \label{Mpi}
\end{eqnarray}
for the $\sigma$ meson pole and the pion poles respectively.  
The bare quark mass parameter $m_0$ is determined so that it generates the physical pions mass in the vacuum, hence it should satisfy near  $q = 0$
the on-the-mass-shell condition $\omega = \sqrt{q^2 + m_\pi^2}$, hence 
\begin{equation}
- m_\pi^2 {\cal F}_{\rm vac} (m_\pi, q )  + {\displaystyle \frac{m_0}{2G M_0} } = 0
\label{pole}
\end{equation}
where $ {\cal F}_{\rm vac} (\omega, q) $ is the vacuum piece of the function ${\cal F} (\omega, q)$: 
\begin{equation}
{\cal F}_{\rm vac} (\omega, q)  =  - \pi  \int_\Lambda \frac{d ^3 p}{(2\pi)^3} \frac{1}{2 E_p 2 E_{p+q}} 
\end{equation}
Subtracting (\ref{pole}) from (\ref{Msigma}) and (\ref{Mpi}) we obtain
\begin{eqnarray}
 {\cal M}_{\sigma} (\omega, q ) & = &  ( - \omega^2 + \bq^2 + 4 M_0^2) {\cal F} (\omega, q) + m_\pi^2 {\cal F}_{\rm vac} (m_\pi, 0)  \\
 {\cal M}_{\pi} (\omega, q) & = &  ( - \omega^2 + \bq^2) {\cal F} (\omega, q) + m_\pi^2 {\cal F}_{\rm vac} (m_\pi, 0) 
\end{eqnarray}
The contribution of these "pole terms"  to the $\omega$-integral is essentially the same as in the chiral limit, except that the masses of mesons are shifted.   

The contribution from the cuts comes from the non-vanishing imaginary part of 
$ {\cal F} ( \omega \pm i \epsilon, q) $, which generates non-vanishing imaginary parts in 
$ {\cal M}_\sigma ( \omega \pm i \epsilon, q) $ by 
\begin{eqnarray}
 {\cal M}_{\sigma}^2 (\omega, q ) & = &  ( - \omega^2 + \bq^2 + 4 M_0^2) {\cal F}_2 (\omega, q) 
%+ {\displaystyle \frac{m_0}{2G M_0} } = 0 
 \label{iMsigma}\\
{\cal M}_{\pi}^2 (\omega, q) & = &  ( - \omega^2 + \bq^2) {\cal F}_2 (\omega, q) 
%+ {\displaystyle \frac{m_0}{2G M_0} } = 0
 \label{iMpi}
\end{eqnarray}
while the real parts of $ {\cal M}_{\pi /\sigma } ( \omega \pm i \epsilon, q) $
%\begin{eqnarray}
%{\cal M}_{\sigma}^1 (\omega, q ) & = &  ( - \omega^2 + \bq^2 + 4 M_0^2) {\cal F}_1 (\omega, q) 
%+ {\displaystyle \frac{m_0}{2G M_0} } 
 %\label{iMsigma}\\
 %{\cal M}_{\pi}^1 (\omega, q) & = &  ( - \omega^2 + \bq^2) {\cal F}_1 (\omega, q) 
%+ {\displaystyle \frac{m_0}{2G M_0} } 
 %\label{iMpi}
%\end{eqnarray}
remain non-vanishing on the cuts, so that we have the phases 
\begin{eqnarray}
\phi_\pi ( \omega, q ) & = & \tan^{-1} \frac{{\cal M}_\pi^2 ( \omega, q )}{{\cal M}_\pi^1 ( \omega, q )} 
= \tan^{-1} \left( \frac{{\cal F}_2}{{\cal F}_1 +  {\displaystyle \frac{m_0}{2G M_0} \frac{1}{ \bq^2- \omega^2} }} \right)
\label{phase_pi} \\
\phi_\sigma ( \omega, q ) & = & \tan^{-1} \frac{ {\cal M}_\sigma^2 ( \omega, q )}{{\cal M}_\sigma^1 ( \omega, q) } 
= \tan^{-1} \left( \frac{{\cal F}_2}{{\cal F}_1 +  {\displaystyle \frac{m_0}{2G M_0} \frac{1}{  \bq^2 + 4 M_0^2 - \omega^2} }} \right)
\label{phase_sigma} 
\end{eqnarray}
shifted slightly from the values in the chiral limit.  
These phases may be interpreted as the phase-shift of the scatterings of quark against quark or anti-quark.\cite{HKZ94,Baym11}

\end{document}